\documentclass[12pt,preprintnumbers,nofootinbib,amsmath,amssymb]{revtex4}
\usepackage{float}
\usepackage{indentfirst}
\usepackage{amsmath,amssymb,epsfig,subfigure}
\usepackage{graphicx}
\usepackage[export]{adjustbox}
\usepackage{color}
\usepackage{multirow}
\usepackage{booktabs}
\usepackage{changes}
\usepackage{footnote}
\usepackage{mathrsfs}

\usepackage{hyperref}
\hypersetup{colorlinks=true,linkcolor=blue,citecolor=magenta}
\begin{document}
\renewcommand{\baselinestretch}{1.15}
\makeatletter
\renewcommand{\thesubfigure}{\alph{subfigure}}
\renewcommand{\@thesubfigure}{(\thesubfigure)\hskip\subfiglabelskip}
\makeatother

\title{Bardeen spacetime with charged scalar field}

\preprint{}

\author{Long-Xing Huang, Shi-Xian Sun, Yong-Qiang Wang \footnote{Corresponding author. E-mail: yqwang@lzu.edu.cn}}

\affiliation{$^{1}$ Lanzhou Center for Theoretical Physics, Key Laboratory of Theoretical Physics of Gansu Province, School of Physical Science and Technology, Lanzhou University, Lanzhou 730000, China\\
 $^{2}$ Institute of Theoretical Physics $\&$ Research Center of Gravitation, Lanzhou University, Lanzhou 730000, China}


\date{\today}

\begin{abstract}
Recently, Ref.~\cite{Wang:2023tdz} investigated the model of Einstein-Bardeen theory coupled to a free complex scalar field. The introduction of the scalar field prevents the formation of the event horizon, and when the magnetic charge exceeds a certain critical value, the frozen Bardeen-boson star can be obtained with the frequency $\omega \rightarrow 0$. In this paper, we extend the investigation of the Einstein-Bardeen model with a charged scalar field and obtain two types of solutions: \textit{the small $q$ solution} and \textit{the large $q$ solution}. Specifically, for the small $q$ solution, we find that there exists a maximum value for the charge $q$, the introduction of the charge makes it possible to obtain solutions for frozen stars without the frequency to be approached to zero. For the large $q$ solution, the charge can tend toward infinity, and as $q \rightarrow \infty$, the large $q$ solution gradually becomes the pure Bardeen solution. Similar to Ref.~\cite{Wang:2023tdz}, the event horizon is not found in our results.
\end{abstract}


\maketitle
\newpage

\section{INTRODUCTION}\label{sec: introduction}
    
     The singularity theorem proved by Penrose and Hawking~\cite{Penrose:1964wq, Hawking:1970zqf, Hawking:1973uf} suggests that when a star meets certain conditions, its collapse process will inevitably lead to the formation of a singularity. Therefore, the existence of singularities appears to be a property inherent to most circumstances in General Relativity. However, the singularity theorem does not deny the existence of singularity-free black holes. In fact, as Hawking and Ellis point out~\cite{Hawking:1973uf}, if there is no strong energy condition or global hyperbolicity, then the singularity theorem will become invalid. 

    Furthermore, the existence of a singularity will cause all laws of physics to fail there and predictability is also lost. Therefore, spacetime singularities are often regarded as a symptom of the limitations of general relativity, similar to the appearance of divergences in some other classical physical theories. It is expected that a suitable theory of quantum gravity will be able to resolve the singularity problem. Unfortunately, we have not yet found a mature and reliable quantum theory. Nevertheless, more phenomenological approaches have attempted to address these singularities in some way. In this context, an important research direction is the so-called regular black holes (RBHs) without singularity. These proposals about RBHs typically postulate the form of a regular black hole metric and then determine what stress tensor would be required to sustain it~\cite{1981NCimL161G, Dymnikova:1992ux, Borde:1994ai, Hayward:2005gi,Lemos:2011dq, Bambi:2013ufa, Simpson:2018tsi, Rodrigues:2018bdc}. Many of these models of the regular black hole can be embedded in models of GR coupled to nonlinear electrodynamics~\cite{Bronnikov:2000vy,Dymnikova:2004zc,Ayon-Beato:2004ywd,Berej:2006cc,Balart:2014jia,Fan:2016rih,Bronnikov:2017sgg,Junior:2023ixh}.

    The earliest ideas about RBHs were proposed in 1966 by Sakharov~\cite{Sakharov:1966aja} and Gliner~\cite{Gliner:1966}, who pointed out that the fundamental singularity could be avoided if the vacuum is replaced by a vacuum-like medium endowed with a de Sitter metric. Later, the first model of the regular black hole was proposed by Bardeen~\cite{Bardeen:1968}. On this basis, several RBH models with similar global spacetime properties to Bardeen spacetime have been studied in 1990s~\cite{Barrabes:1995nk,Mars:1996khm,Frolov:1988vj}, and these models are also often uniformly called ``Bardeen black holes”~\cite{Borde:1996df}. It is worth noting that the Bardeen black hole, when initially proposed, is not the exact solution of Einstein's equations and has no known physically reasonable source. The underlying physical motivation for this choice was proposed two decades ago by Ayon-Beato and Garcia, who reinterpreted the Bardeen model in terms of the gravitational field possessed by magnetic monopoles~\cite{Ayon-Beato:1998hmi,Ayon-Beato:2000mjt}, namely, the Bardeen solution is a magnetic solution to Einstein's field equations coupled to a nonlinear electromagnetic field. Apart from the Bardeen black hole, at present, there is a wealth of literature on the study of RBHs, garnering growing attention and undergoing continuous development (see review articles~\cite{Ansoldi:2008jw,Lan:2023cvz,Torres:2022twv}).

    However, the physical sources of RBHs have only considered the nonlinear electromagnetic field in most of the literature, with no consideration of other types of matter fields. Recently, Ref.~\cite{Wang:2023tdz} studied the model of the Einstein-Bardeen theory minimally coupled to a free complex scalar field and then found that after introducing a scalar field into the Bardeen spacetime, the magnetic charge of the nonlinear electromagnetic field will have an upper limit. Within this upper limit, the black hole solution is not found. In particular, when the frequency approaches zero, the scalar field of this model is almost concentrated inside a certain critical horizon $r_{cH}$  and the metric function inside the region $r\leq r_{cH}$ is nearly zero, while the other metric function $g^{rr}$ is nearly zero only at $r_{cH}$. Notice that since $g^{rr}$ is only nearly zero but not zero at $r_{cH}$, the surface represented by $r=r_{cH}$ is actually not an event horizon, and this solution is not a black hole but a frozen star. The term ``frozen" is used because only $-g_{tt}$ is nearly zero within $r_{cH}$, leading objects within $r_{cH}$ taking a very long time to move a small distance from the perspective of an observer at infinity, as if they are frozen in place (see Ref.~\cite{Oppenheimer:1939ue,zeldovichbookorpaper,Ruffini:1971bza} for the early study on the frozen star). In fact, these solutions are essentially also a type of boson stars with a magnetic charge, in which boson stars are a theoretical concept in astrophysics characterized by their ability to remain stable and avoid gravitational collapse~\cite{Wheeler:1955zz,Power:1957zz,Kaup:1968zz,Ruffini:1969qy} (see Refs.~\cite{Schunck:2003kk,Liebling:2012fv} for a review). After Ref.~\cite{Wang:2023tdz}, the model of Einstein-Klein-Gordon theory coupled to another nonlinear electromagnetic field proposed by Hayward~\cite{Yue:2023sep} and the model of Bardeen spacetime with the free Dirac field~\cite{Huang:2023fnt}  have also been studied. Same as ~\cite{Wang:2023tdz}, in these two works, the event horizon and black hole solution are not found after the matter field is introduced, and when the frequency of the matter field approaches zero, the frozen star also can be obtained.
    
  It is worth noting that the matter fields considered in Ref.~\cite{Wang:2023tdz,Huang:2023fnt,Yue:2023sep} are all uncharged, so one may wonder, in addition to these free matter fields, whether the solution of the frozen star can still be found after introducing charged matter fields. Considering that, in gravitational physics, one of the simplest types of ``charged matter" that is often considered is the charged scalar field. Thus, to answer this question preliminary, in this paper, we further study the influence of the charged scalar field on the Bardeen spacetime, i.e., the model of Einstein-Bardeen theory coupled with a charged scalar field. By solving the Einstein equations, we find that after introducing the linear electromagnetic field (i.e., $U(1)$ gauge field), two types of solutions can be obtained - the small $q$ solution and the large $q$ solution. For the small $q$ solution, we find that there is a maximum value for the electric charge $q$, and after introducing $U(1)$ gauge field, the frozen star can be obtained without the frequency very close to zero. For the large $q$ solution, the electric charge can tend to infinity, and as the $q$ increases, the large $q$ solution gradually transforms into the pure Bardeen solution. In addition, similar to Ref.~\cite{Wang:2023tdz,Huang:2023fnt,Yue:2023sep}, there is no event horizon and no black hole solution in our results.

    The paper is organized as follows. In Sec.~\ref{sec: model}, we provide an introduction to the model of Einstein-Bardeen theory coupled to a charged complex scalar field. Numerical results of this model and analysis of its physical properties are present in Sec.~\ref{sec: results}. The conclusion and discussion are given in Sec.~\ref{sec: conclusion}.

\section{MODEL}\label{sec: model}
 In this section, we wish to provide a concise introduction to the theoretical framework encompassing the Einstein-nonlinear electrodynamics model, coupled with a charged complex scalar field $\psi$, described by the following action (we use units with $c = 1 = \hbar$)
\begin{equation}
     S=\int d^4 x \sqrt{-g}\left [\frac{R}{16\pi G}+\mathcal{L}^{(1)}+\mathcal{L}^{(2)}+\mathcal{L}^{(3)}\right],
         \label{eq:action}
\end{equation}
where $\mathcal{L}^{(1)}$, $\mathcal{L}^{(2)}$ and $\mathcal{L}^{(3)}$ are the Lagrangians of the nonlinear electromagnetic field~\cite{Ayon-Beato:2000mjt}, the linear electromagnetic field ($U(1)$ gauge field) and the scalar field, respectively,
\begin{equation}
     \mathcal{L}^{(1)}=\frac{3}{2s}\left( \frac{\sqrt{p^2E^{\mu\nu}E_{\mu\nu}}}{\sqrt{2}+\sqrt{p^2 E^{\mu\nu}E_{\mu\nu}}} \right)^{(5/2)},
\end{equation}
\begin{equation}
     \mathcal{L}^{(2)}=-\frac{1}{4}F^{\mu \nu} F_{\mu \nu},
\end{equation}
\begin{equation}
     \mathcal{L}^{(3)}=-\left(D^{\mu}\psi \right)^*\left(D_{\mu}\psi \right)-\mu^2\psi\psi^*,
\end{equation}
with $R$ the Ricci curvature, $G$ Newton’s constant. $E^{\mu \nu}$ and $F^{\mu \nu}$ are the electromagnetic field strength of the nonlinear electromagnetic field $A_{\mu}$ and the linear electromagnetic field $B_{\mu}$, respectively, given in terms of the potential 1-form~\cite{Ayon-Beato:2000mjt,Jetzer:1990wr,Fulling,BD} as:
\begin{equation}
       E_{\mu \nu}:=\partial_\mu A_\nu-\partial_\nu A_\mu,
\end{equation}
\begin{equation}
   F^{\mu \nu}:=\partial_{\mu} B_{\nu}-\partial_{\nu} B_{\mu}.
\end{equation}
and $D_{\mu}$ denotes the gauge invariant covariant derivative:
 \begin{equation}
     D_{\mu}:=\nabla_{\mu}+iq B_{\mu},
 \end{equation}
where $\nabla$ stands for the spacetime covariant derivative and the constant $q$ is the scalar field electric charge. The constants $s$, $p$, and $\mu$ are three independent parameters, where $p$ represents the magnetic charge and $\mu$ is the scalar field mass.

The action (\ref{eq:action}) is invariant under local $U (1)$ gauge transformations $\psi\rightarrow e^{iq\alpha(x^{\theta })}\psi$, where $\alpha(x^{\theta })$ a local gauge function which depends on the spacetime coordinates. As a result, the system possesses a conserved Noether density current, which acts as a source for the electromagnetic field $B^{\mu}$ and is given by:
\begin{equation}
    j^{\nu}=i g^{\mu\nu} q\left[ \psi^*D_{\nu}\psi-\psi(D_{\nu}\psi^*)\right].
\end{equation}
This implies the existence of a total electric charge:
\begin{equation}
    Q=\int_{\mathcal{S}} j^{\nu}n_{\nu}dV, \label{eq:charge}
\end{equation}
here, $\mathcal{S}$ is a spacelike hypersurface (with the volume element $dV$), and $n_{\nu}$ is the unit normal vector of $\mathcal{S}$. The total scalar particle number $N$~\cite{Jetzer:1989av} can be obtained through the total charge $Q$, where
\begin{equation}
    N=\frac{Q}{q}.
\end{equation}

By varying the action (\ref{eq:action}) with respect to the metric $g_{\mu \nu}$, electromagnetic field $A_{\mu}$, $B_{\mu}$ and scalar field $\psi$, respectively, we can obtain the following equations of motion
\begin{equation}
    R_{\mu \nu} - \frac{1}{2}g_{\mu \nu}R - 8\pi G (T^{(1)}_{\mu \nu}+T^{(2)}_{\mu \nu}+T^{(3)}_{\mu \nu})=0,
    			\label{eq:einstein}
\end{equation}
\begin{equation}
    \nabla_{\mu}\left(\frac{\partial \mathcal{L}^{(1)}}{\partial \mathcal{E}} E^{\mu \nu}\right)=0,
    			\label{eq:equationBardeen}
\end{equation}
\begin{equation}
    \nabla_{\mu}F^{\mu \nu}=-j^{\nu},
    			\label{eq:equationMaxwellF}
\end{equation}
\begin{equation}
    D^{\mu}D_{\mu}\psi-\mu^2\psi=0,
    			\label{eq:equationKG}
\end{equation}
with the stress-energy tensor
\begin{equation}
    T^{(1)}_{\mu \nu}=-\frac{\partial \mathcal{L}^{(1)}}{\partial \mathcal{E}} E_{\mu \lambda} E_\mu{ }^\lambda+g_{\mu \nu} \mathcal{L}^{(1)},
\end{equation}
\begin{equation}
    T^{(2)}_{\mu \nu}=-\frac{\partial \mathcal{L}^{(2)}}{\partial \mathcal{F}} F_{\mu \lambda} F_\mu{ }^\lambda+g_{\mu \nu} \mathcal{L}^{(2)},
\end{equation}
\begin{equation}
    T^{(3)}_{\mu \nu}=\frac{1}{2}\left[\left(D_\mu \psi\right)^*\left(D_\nu \psi\right)+\left(D_\nu \psi\right)\left(D_\mu \psi\right)^*-g_{\mu \nu}\left(\left(D_\lambda \psi\right)\left(D^\lambda \psi\right)^*+m^2|\psi|^2\right)\right],
\end{equation}
where $\mathcal{E}=\frac{1}{4}E^{\mu \nu} E_{\mu \nu}$ and $\mathcal{F}=\frac{1}{4}F^{\mu \nu} F_{\mu \nu}$.

Restricting to static, spherically-symmetric solutions of the field equations, we employ the following ansatz
\begin{equation}
    d s^2=-n(r) \sigma^2(r) d t^2+\frac{d r^2}{n(r)}+r^2\left(d \theta^2+\sin ^2 \theta d \varphi^2\right),
    \label{eq:metric}
\end{equation}
here, $n(r)$ and $\sigma(r)$ are functions of the radial coordinate $r$ only. Then, for the electromagnetic fields~\cite{Jetzer:1990wr,Kain:2021bwd} and the scalar field~\cite{Derrick:1964ww,Wyman:1981bd,Adib:2002fd,Schunck:2003kk}, we assume
\begin{equation}
    A_{\mu}dx^{\mu}=p\cos(\theta)d\varphi,\quad B_{\mu}dx^{\mu}=V(r)dt,\quad \psi=\phi(r) e^{-i\omega t}, 
    \label{eq:ansatz}
\end{equation}
where the function $V(r)$ corresponding to the electric potential and the function $\phi(r)$ are two real radial functions, and $\omega>0$ is the real constant corresponding to the frequency of oscillation of the scalar field. 

Using the metric (\ref{eq:metric}) and the ansatz (\ref{eq:ansatz} ) for the scalar field and electromagnetic fields into the equations (\ref{eq:einstein}-\ref{eq:equationKG}), we can get the following equations about $n(r)$, $\sigma(r)$ $V(r)$, and $\phi(r)$
\begin{align}
    \notag &n^{\prime}+n\left(\frac{1}{r}+8\pi Gr\phi^{\prime 2}\right)+\frac{8\pi Gr \phi^{2}}{n\sigma^2}\left(\omega^2-2 q \omega V+q^2 V^2\right)\\ 
    &+\frac{4\pi Gr V^2}{\sigma}+8\pi Gr\mu^2\phi^{ 2}+\frac{12\pi Gr p^5}{(p^2+r^2)^{5/2}s}-\frac{1}{r}=0,\label{eq:n}
\end{align}
\begin{equation}
    \sigma^2-8\pi Gr\phi^{\prime 2}\sigma^2-\frac{8 \pi G r \phi^2}{n^2\sigma} \left(q^2V^2-2q\omega V+\omega^2\right),
\end{equation}
\begin{equation}
V^{\prime \prime}+V^{\prime}\left(\frac{2}{r} -\frac{ \sigma^{\prime}}{\sigma}\right)+\frac{2q\phi^2}{n}\left(q V-\omega \right)=0,
\end{equation}
 \begin{equation}
\phi^{\prime \prime}+\phi^{\prime}\left( \frac{n^{\prime}}{n}+\frac{\sigma^{\prime}}{\sigma}+\frac{2}{r}\right)+\left(\frac{q^2 V^2}{n\sigma^2}+\frac{\omega^2}{n\sigma^2}-\frac{2q\omega V}{n\sigma^2}  -\mu^2\right)\frac{\phi}{n}=0.\label{eq:phi}
\end{equation}
where the prime denotes the differentiation with respect to $r$. And form equations (\ref{eq:charge}), (\ref{eq:metric}), (\ref{eq:ansatz}), the total electric charge is:
\begin{equation}
    Q= 8\pi G q\int^{\infty}_0   \frac{qV-\omega}{n\sigma} r^2\phi^2.
\end{equation}

Before numerically solving these coupled equations of motion (\ref{eq:n}-\ref{eq:phi}), we need to impose suitable boundary conditions for $n(r)$, $\sigma(r)$, $\phi(r)$, $V(r)$. Considering the assumption of regularity and asymptotic flatness, the metric functions $n(r)$ and $\sigma(r)$ satisfy:
\begin{equation}
    n(0)=1,\quad \sigma(0)=\sigma_0,\quad n(\infty)=1-\frac{2GM}{r},\quad \sigma(\infty)=1,
\end{equation}
For the constants, namely $\sigma_0$ and the ADM mass $M$ of the solution, their values can be determined by solving the system of ordinary differential equations. Additionally, for the complex scalar field and the $U(1)$ gauge field $B_{\mu}$, we require the following boundary conditions
\begin{equation}
    \left.\frac{d \phi}{d r}\right|_{r=0}=0,\quad \phi(\infty)=0,\quad \left.\frac{d V}{d r}\right|_{r=0}=0,\quad V(\infty)=0.
\end{equation}
It is worthwhile to investigate two special solution cases of these equations of motion (\ref{eq:n}-\ref{eq:phi}). First, when $p = 0$ and the scalar field does not vanish, action (\ref{eq:action}) reduces to Einstein-Maxwell-scalar theory which can describe charged boson star~\cite{Jetzer:1989av} (Further, if $p=q=0$, this model describes pure boson star). Then, in the case of vanishing complex scalar field with $q=0$ and $p\neq0$, the model is the Bardeen theory and the associated solution describes both spherical solitons or black holes, also referred to as the Bardeen spacetime. The metric is the following form
\begin{equation}
d s^2=-f(r) d t^2+f(r)^{-1} d r^2+r^2\left(d \theta^2+\sin ^2(\theta) d \varphi^2\right),
\end{equation}
here, 
\begin{equation}
    f(r)=1-\frac{p^3r^2}{s(r^2+p^2)^{3/2}}.
\end{equation}
The asymptotic behavior for this metric function $f(r)$ at infinity is given by
\begin{equation}
    f(r)=1-2\frac{p^3}{2s}/r+O(1/r^3).
\end{equation}
From the term $1/r$, one can deduce the ADM mass $M$ of this configuration as $M = p^3/2s$~\cite{Ayon-Beato:1998hmi}. Furthermore, the function $f(r)$ demonstrates a local minimum value at $ r=\sqrt{2}p$. In case where $q<3^{3/4}\sqrt{\frac{s}{2}}$, the solutions without event horizons are present. When $q=3^{3/4}\sqrt{\frac{s}{2}}$, degenerate horizons are observed, and for $q>3^{3/4}\sqrt{\frac{s}{2}}$, two distinct horizons exist.
\section{NUMERICAL RESULTS}\label{sec: results}
To facilitate numerical computations, we employ the following dimensionless  variables introduced by using natural units set by $\mu$ and $G$:
\begin{equation}
    r\rightarrow r\mu,\quad \omega\rightarrow\frac{\omega}{\mu}, \quad  \phi\rightarrow\frac{\sqrt{4\pi}}{M_{Pl}}\phi,\quad V\rightarrow\frac{\sqrt{4\pi}}{M_{Pl}}V,                          
    \end{equation}
where $M_{Pl}=1/\sqrt{G}$ denote the Planck mass, Consequently, the dependence on both $G$ and $\mu$ disappears from the equations. Furthermore, the total electric charge and the ADM matter are also can be expressed in units set by $\mu$ and $G$
(Note that, to simplify the output, without loss of generality, we set $s = 0.2$ in the following results). 

We also introduce a new radial coordinate $x = r/(r + 1)$, which maps the semi-infinite region $[0, \infty)$ onto the unit interval $[0, 1]$. In addition, the solutions are found by using the finite element method, and the number of grid points in the integration region $[0, 1]$ is $1000$. The iterative method we use is the Newton-Raphson method, and the relative error for the numerical solutions in this paper is estimated to be below $10^{-5}$.

In our numerical results, we find that the magnetic charge $p$ can only be less than $3^3/4\sqrt{s/2}=3^3/4\sqrt{0.1}$ (i.e., $p<3^3/4\sqrt{0.1}$). Within this range, two different types of solutions emerge. Based on the electric charge $q$, we refer to these two types of solutions as the ``small $q$ solution" and the ``large $q$ solution". The small $q$ solution is a set of solutions with relatively small electric charges, which have an upper limit on $q$. In contrast, the large $q$ solution is a set of solutions with relatively large electric charges, which can tend to infinity but have a lower limit. We present the ranges of $q$ for the small $q$ solution and the large $q$ solution in the second and fourth columns of Tab. \ref{tab:rangeofq}, and the fourth column of Tab. \ref{tab:rangeofq} corresponds to the critical charge $q_c$ for the small $q$ solution. 

  	\begin{table}[!htbp]     
    \scriptsize 
	\centering 
	\begin{tabular}{c||cc||c}
		\hline
		&\multicolumn{2}{c||}{Small $q$ solution} &\multicolumn{1}{c}{Large $q$ solution}  \\
		\hline
		 $p$ & $q_s$ & $q_c$ & $q_l$ \\  \hline  \hline
		$0$ & $0\sim1$ & $1$ & $-$  \\  
		$0.3$ & $0\sim1.077$ & $1.003$ & $1.005\sim\infty$  \\ 
		$0.5$ & $0\sim1.543$ & $1.085$ & $1.101\sim\infty$  \\ 
  	$0.65$ & $0\sim1.425$ & $1.331$ & $1.347\sim\infty$  \\ 
		$0.71$ & $0\sim1.422$ & $1.413$ & $1.424\sim\infty$  \\ 
		\hline
	\end{tabular}
 	\caption{$1^{st}$ column: different magnetic charge $p$; $2^{nd}$ and $4^{th}$ column: the range of electric charge for small $q$ solution ($q_s$) and large $q$ solution ($q_l$); $3^{rd}$ column: critical charge $q_c$ of small $q$ solution.}
	\label{tab:rangeofq}
\end{table}

\subsection{Small $q$ solution }

\begin{figure}[!htbp]
		\begin{center} 
     		\subfigure{  
			\includegraphics[height=.28\textheight,width=.30\textheight, angle =0]{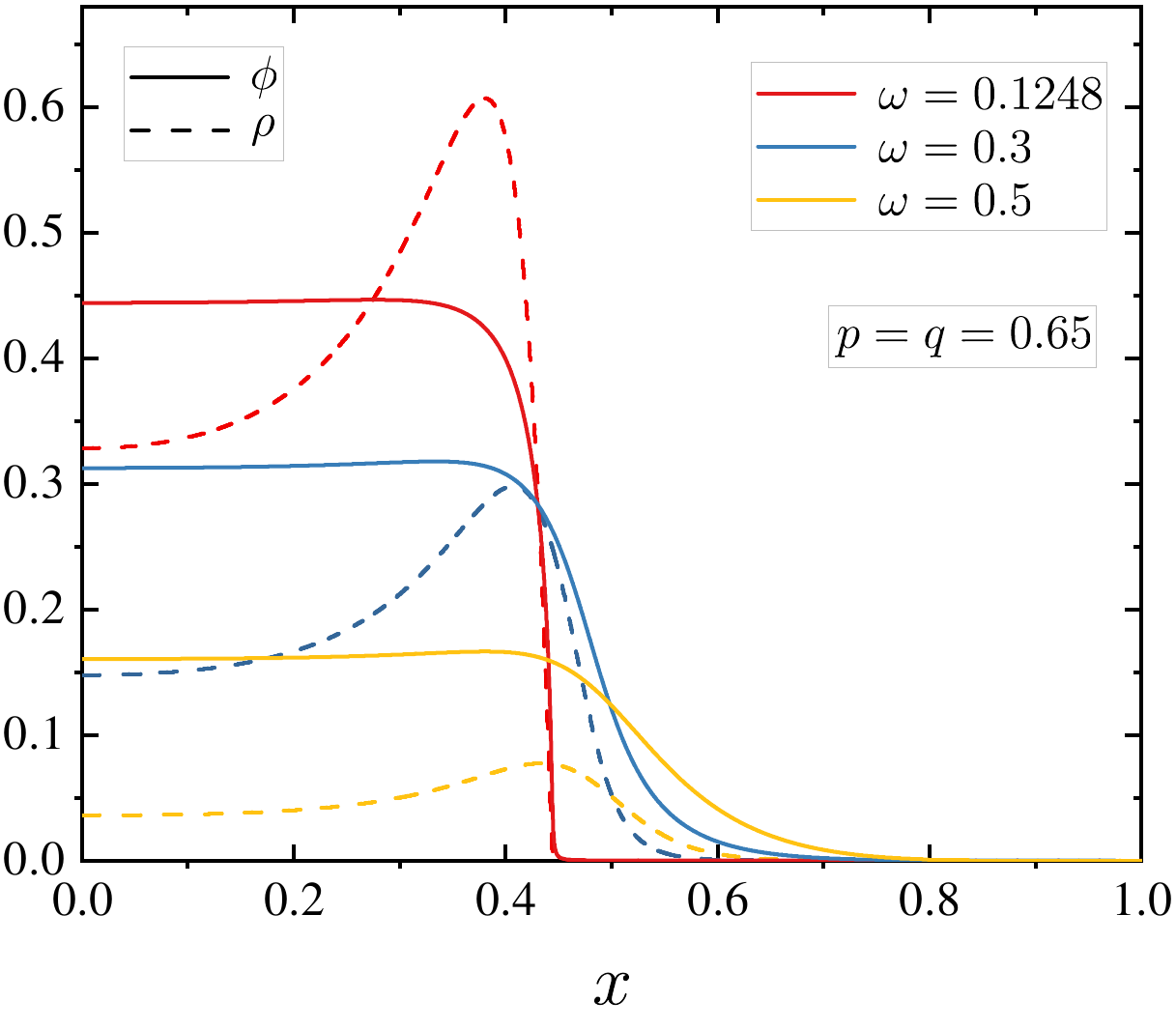}
			\label{fig:funcphi6565}
		}
  		\subfigure{  
			\includegraphics[height=.28\textheight,width=.30\textheight, angle =0]{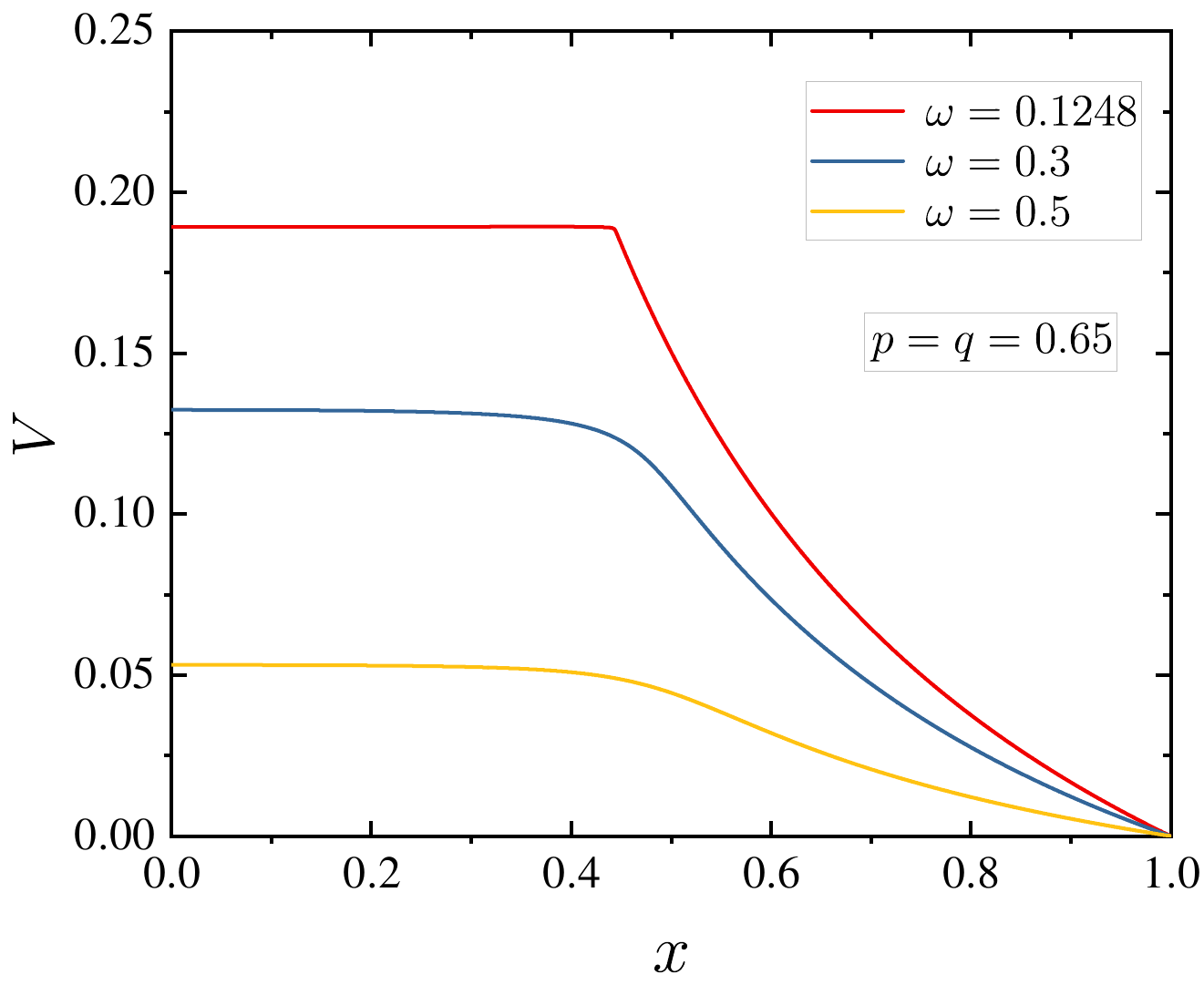}
			\label{fig:funca6565}
		}	
  		\end{center}	
		\caption{The radial distribution of the scalar field function $\phi$ (the solid line in the left panel),  the energy density $\rho$ of the scalar field (the dashed line in the left panel), and the electromagnetic field $V$ (the right panel). All solutions have $p = q = 0.65$.}
		\label{fig:funmatter6565}		
		\end{figure}

As previously mentioned, when the electric charge is relatively small, there exists a solution which we refer to as the small $q$ solution, and this solution has an upper limit for the electric charge. Taking $p = q = 0.65$ as an example, we present the distribution of the scalar field $\phi$, the energy density $\rho=-T_0^{(3) 0}$ and the electric potential $V$ as a function of $x$ coordinate in Fig.~\ref{fig:funmatter6565}. On the one hand, it can be observed that as the frequency increases, the scalar field function $\phi$, the energy density $\rho$, and the electric potential $V$ gradually decrease. In our results, it can be proven that finally, when the frequency increases to a certain value, $\phi$, $\rho$, and $V$ vanish. At this time, the obtained solution is no longer a small $q$ solution but a pure Bardeen solution. After this, regardless of how the frequency changes, the obtained solution remains a pure Bardeen solution. On the other hand, as the frequency decreases to the minimum value $0.1248$, both the scalar field function and energy density form a steep, wall-like shape near a certain position $x_{cH}$ (i.e., the position $x_{cH}$, or equivalently the position $r_{cH}$, is ``critical position", with $r_{cH}=x_{cH}/(1-x_{cH})$), and the scalar field almost concentrates within $x_{cH}$ - see the red line in the left panel.

\begin{figure}[!htbp]
		\begin{center}
    		\subfigure{  
			\includegraphics[height=.28\textheight,width=.30\textheight, angle =0]{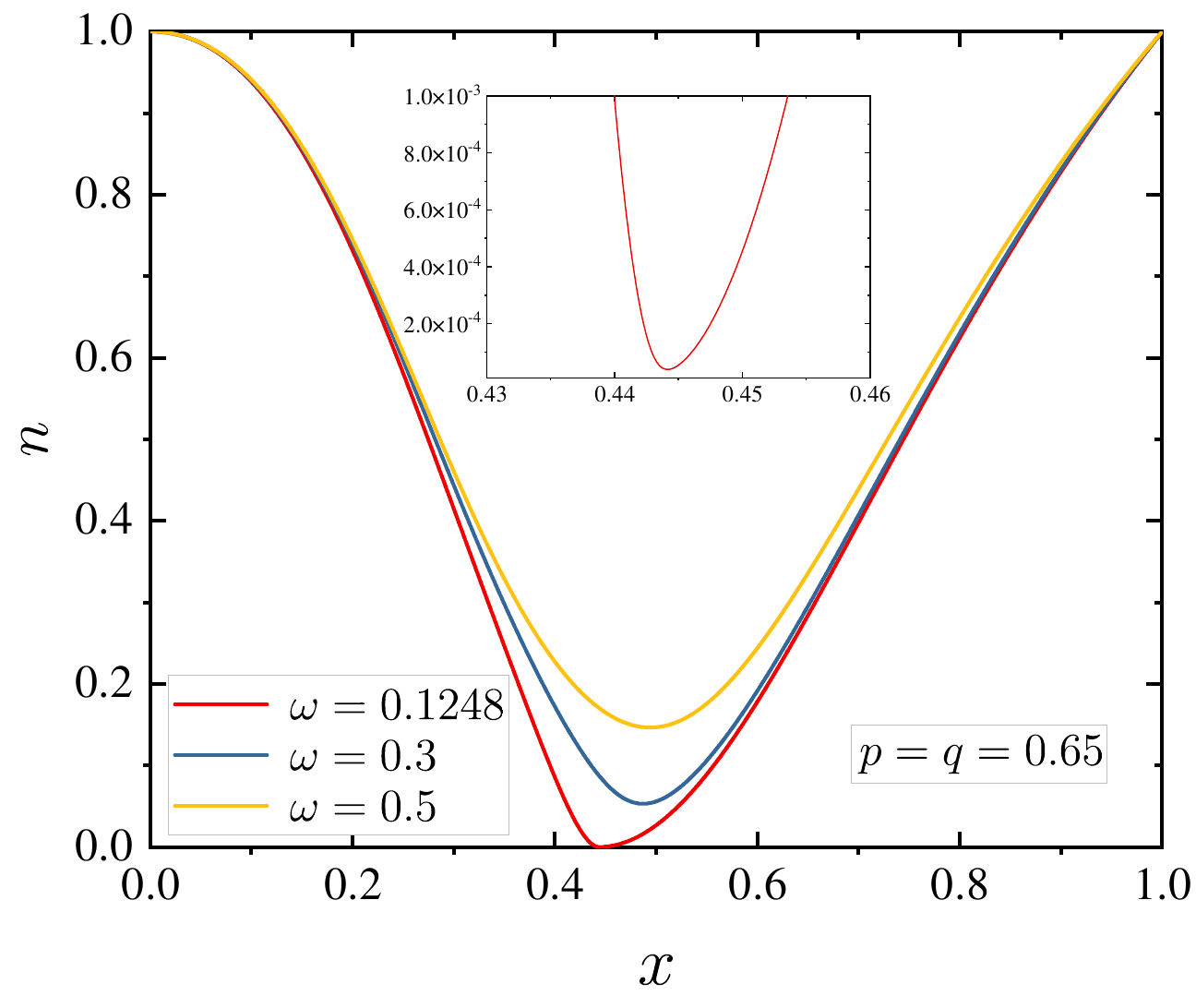}
			\label{fig:funcn6565}
		}
    		\subfigure{  
			\includegraphics[height=.28\textheight,width=.30\textheight, angle =0]{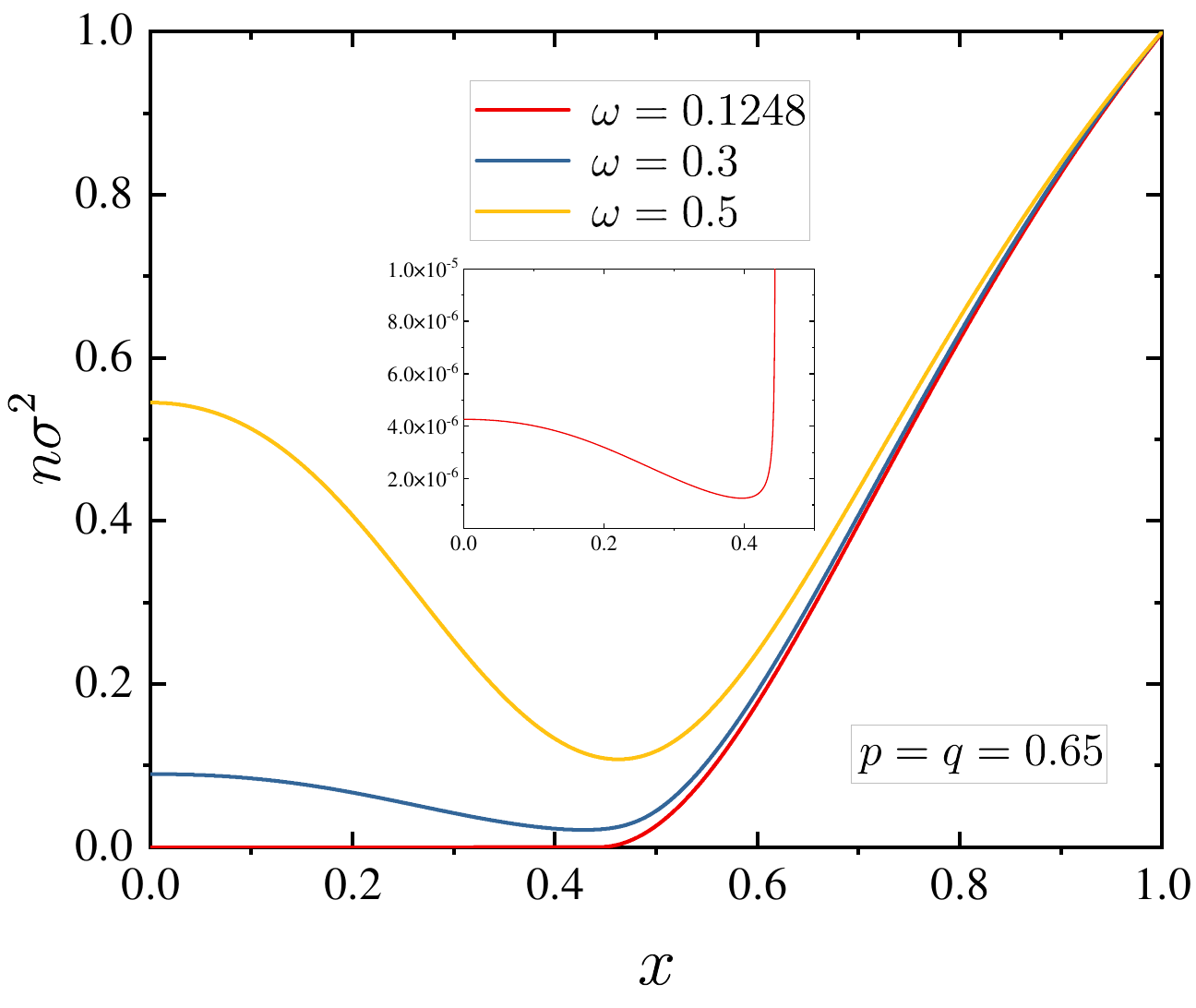}
			\label{fig:funcno6565}
		}
  		\end{center}	
		\caption{The radial distribution of the metric field functions with $p = q = 0.56$.}
		\label{fig:funcmetirc6565}		
		\end{figure}

\begin{figure}[!htbp]
		\begin{center}
		\subfigure{ 
			\includegraphics[height=.28\textheight,width=.30\textheight, angle =0]{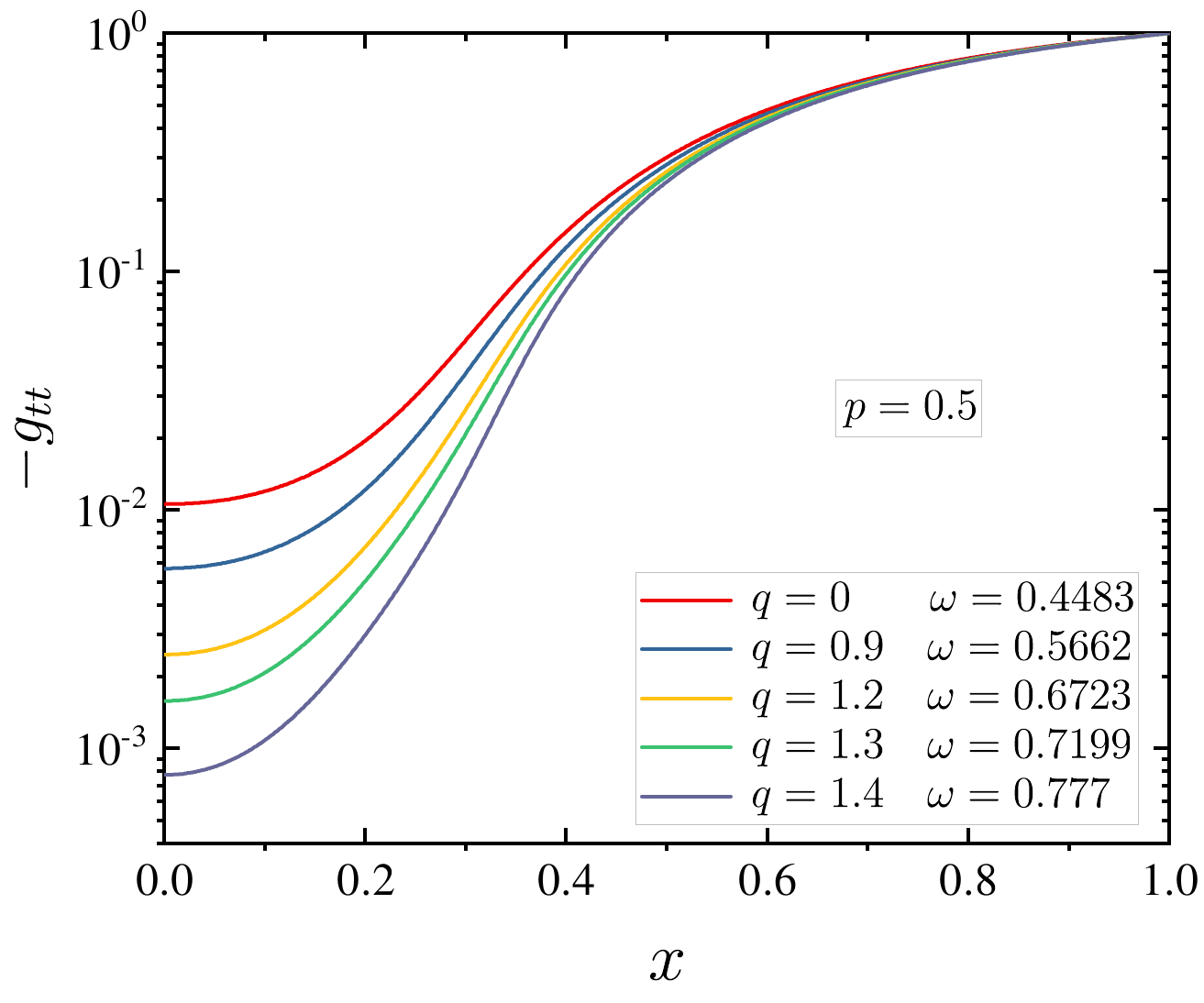}
			\label{fig:funcnochange50}
		}	 
  		\subfigure{  
			\includegraphics[height=.28\textheight,width=.30\textheight, angle =0]{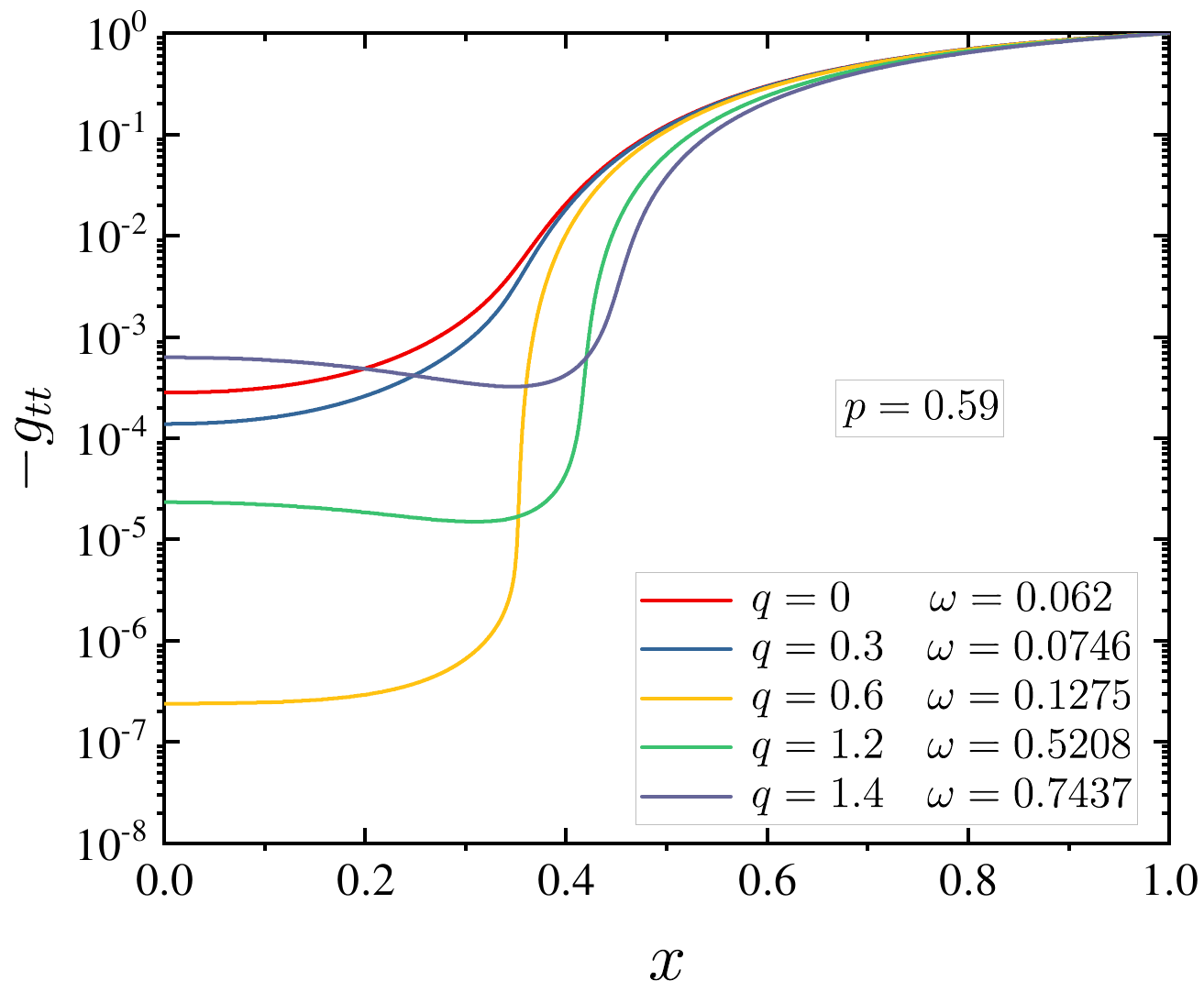}
			\label{fig:funcnochange59}
		}
  \quad
  		\subfigure{ 
			\includegraphics[height=.28\textheight,width=.30\textheight, angle =0]{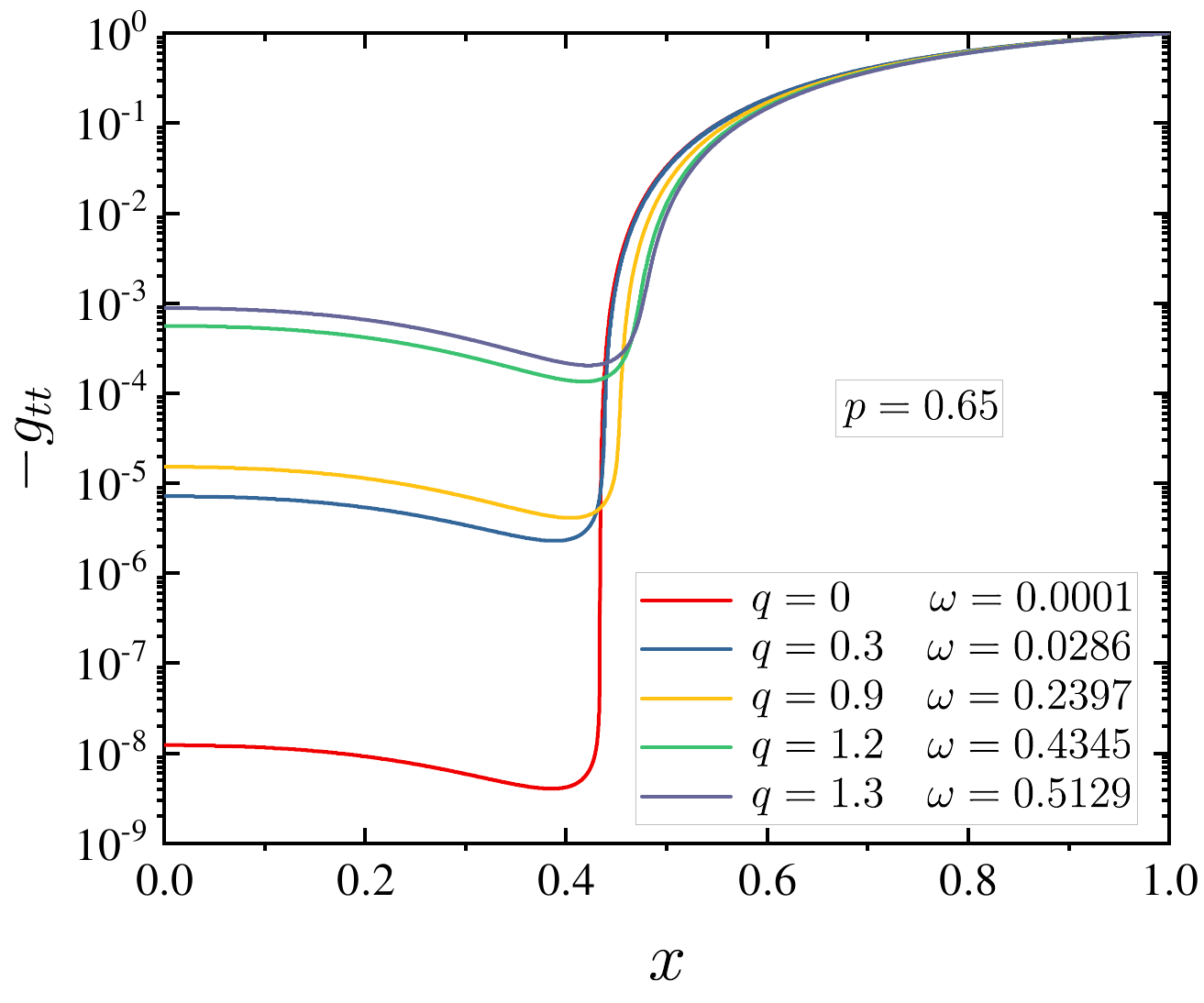}
			\label{fig:funcnochange65}
		}	 
  		\end{center}	
		\caption{ The metric field function $-g_{tt}$ as a function of $x$ with $p=0.5,0.59,0.65$ for small $q$ solution with several electric charge.}
	\label{fig:funcnochange}	
		\end{figure}

For the same magnetic charge and electric charge, it can be seen in Fig.~\ref{fig:funcmetirc6565} that as the frequency decreases, the minimum values of the metric field function $g^{rr}=n(r)$ and $-g_{tt}=n(r)\sigma(r)^2$ are gradually decreasing. When the frequency $\omega$ reaches $0.1248$, the location of the minimum value of the function $g^{rr}$ and $-g_{tt}$ is the critical position $x_{cH}$, and both the functions $g^{rr}$ and $-g_{tt}$ is nearly zero at $x_{cH}$ (below $10^{-4}$). In particular, for the function $-g_{tt}$, it is also close to zero within $x_{cH}$, and the order of magnitude of $-g_{tt}$ within $x_{cH}$ is almost equal to the order of magnitude of the minimum value of $-g_{tt}$ (i.e., $-g_{tt}(x_{cH})$). In fact, with the improvement in computational accuracy, the minimum value of $-g_{tt}$ and $g^{rr}$ can be made even closer to zero\footnote{For the small $q$ solutions with $p=q=0.65$, $\omega=0.1248$ is almost a minimum value of frequency. With further improvements in computational accuracy, we find that the frequency can increase by more digits (e.g., $\omega=0.12479$), and the minimum value of the metric function can be closer to zero.}. Nevertheless, it is worth noting that in our calculation results, they only get closer and closer to zero, but never actually equal zero. Therefore, the surface with the constant radial variable $x=x_{cH}$ (or equivalently, $r=r_{cH}$) is not an event horizon and this solution is not the black hole. But considering that these properties of the metric function at the  position $x_{cH}$ are very similar to those at the event horizon, we can refer to the surface represented by $x=x_{cH}$ as ``critical horizon". Meanwhile, since the scalar field
of the solution with $\omega=0.1248$ converges at the critical horizon while rapidly decaying beyond that radius, and only $-g_{tt}$ is nearly zero (below $10^{-5}$) inside the region $x\leq x_{cH}$, the solution with $\omega=0.1248$ is also the frozen star solution\footnote{Since the location of the minimum value of the function $g_{tt}$ is the critical position and the order of magnitude of the function $-g_{tt}$ inside the region $x\leq x_{cH}$ is almost unchanged in our results - see Fig.~\ref{fig:funcmetirc6565} and Fig.~\ref{fig:funcnochange} as illustrative cases, the function $-g_{tt}$ being close to zero within $x_{cH}$ is almost equivalent to its minimum value $-g_{tt(min)}$ being close to zero. Thus, for the convenience of explanation and considering numerical errors and the requirements for computational accuracy, without loss of generality, we can use a criterion that solutions with $-g_{tt(min)}<10^{-5}$ are considered frozen stars in this paper.}. Unlike the previous consideration of a free scalar in Bardeen spacetime~\cite{Wang:2023tdz}, in this case, obtaining the solution of the frozen star does not require the frequency to be very close to zero.

\begin{figure}[!htbp]
		\begin{center}
		\subfigure{ 
			\includegraphics[height=.28\textheight,width=.30\textheight, angle =0]{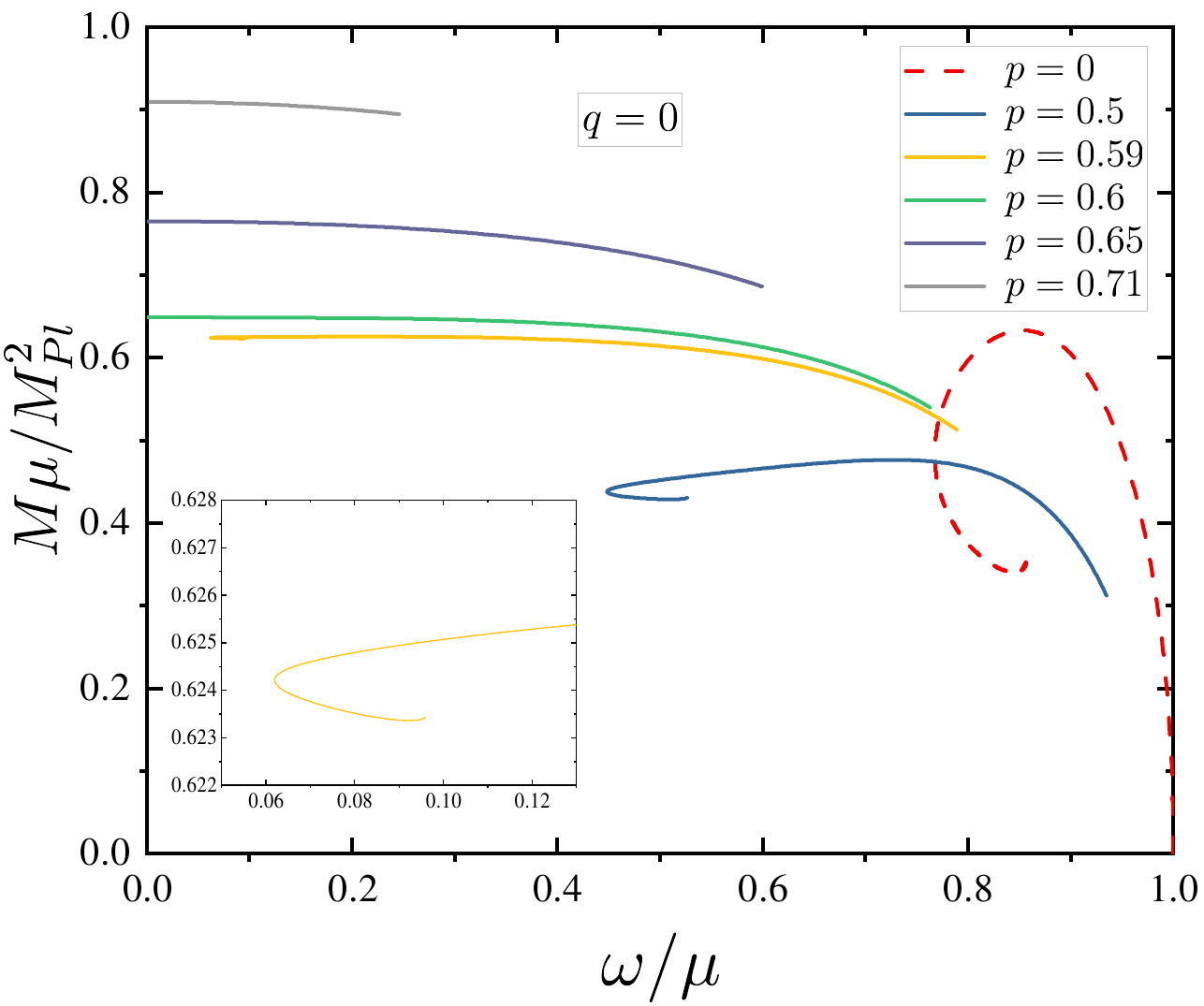}
			\label{fig:cbbsmatterfor02andall0}
		}	 
  		\subfigure{  
			\includegraphics[height=.28\textheight,width=.30\textheight, angle =0]{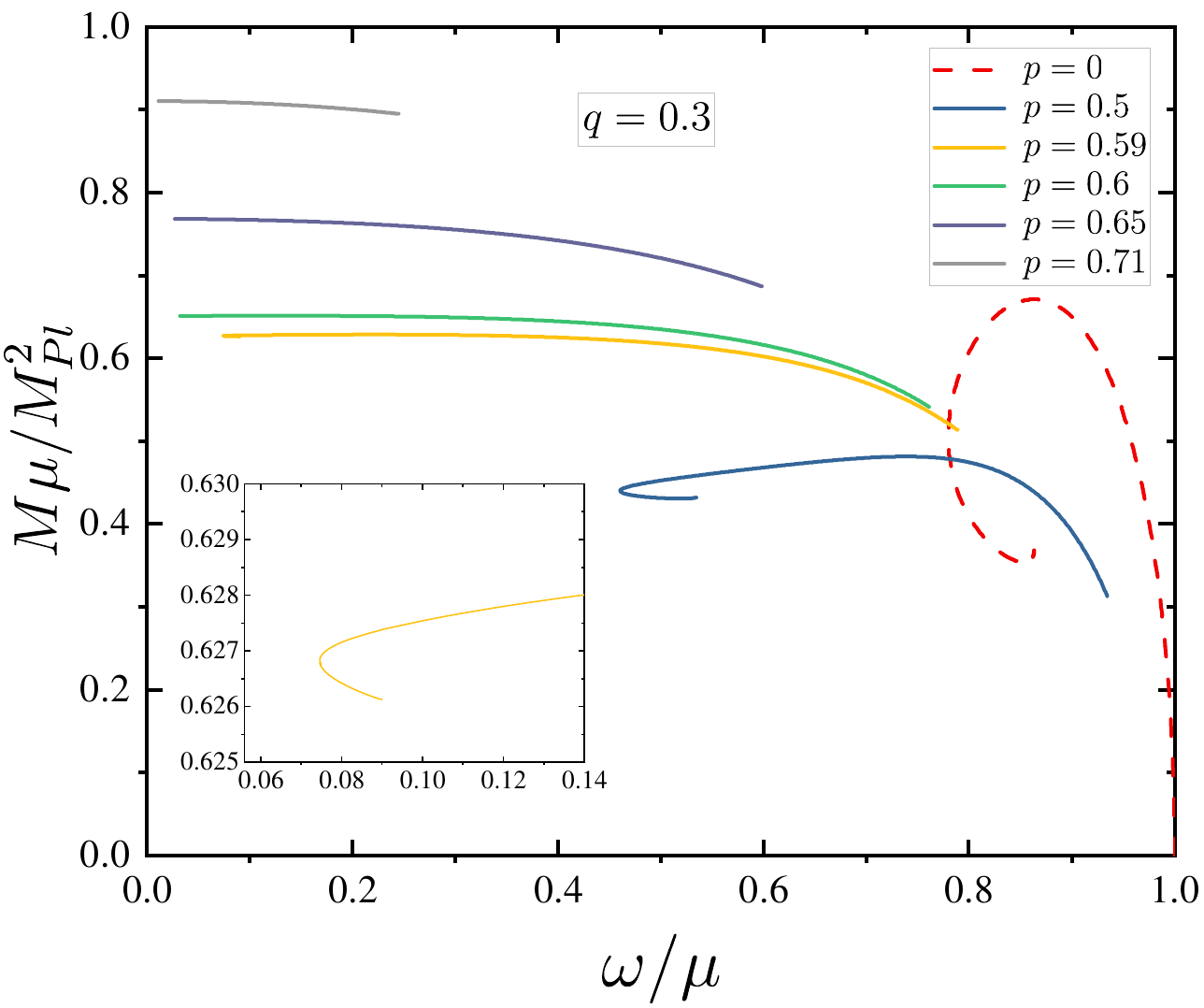}
			\label{fig:cbbsmatterfor02andall03}
		}	
  \quad
    		\subfigure{  
			\includegraphics[height=.28\textheight,width=.30\textheight, angle =0]{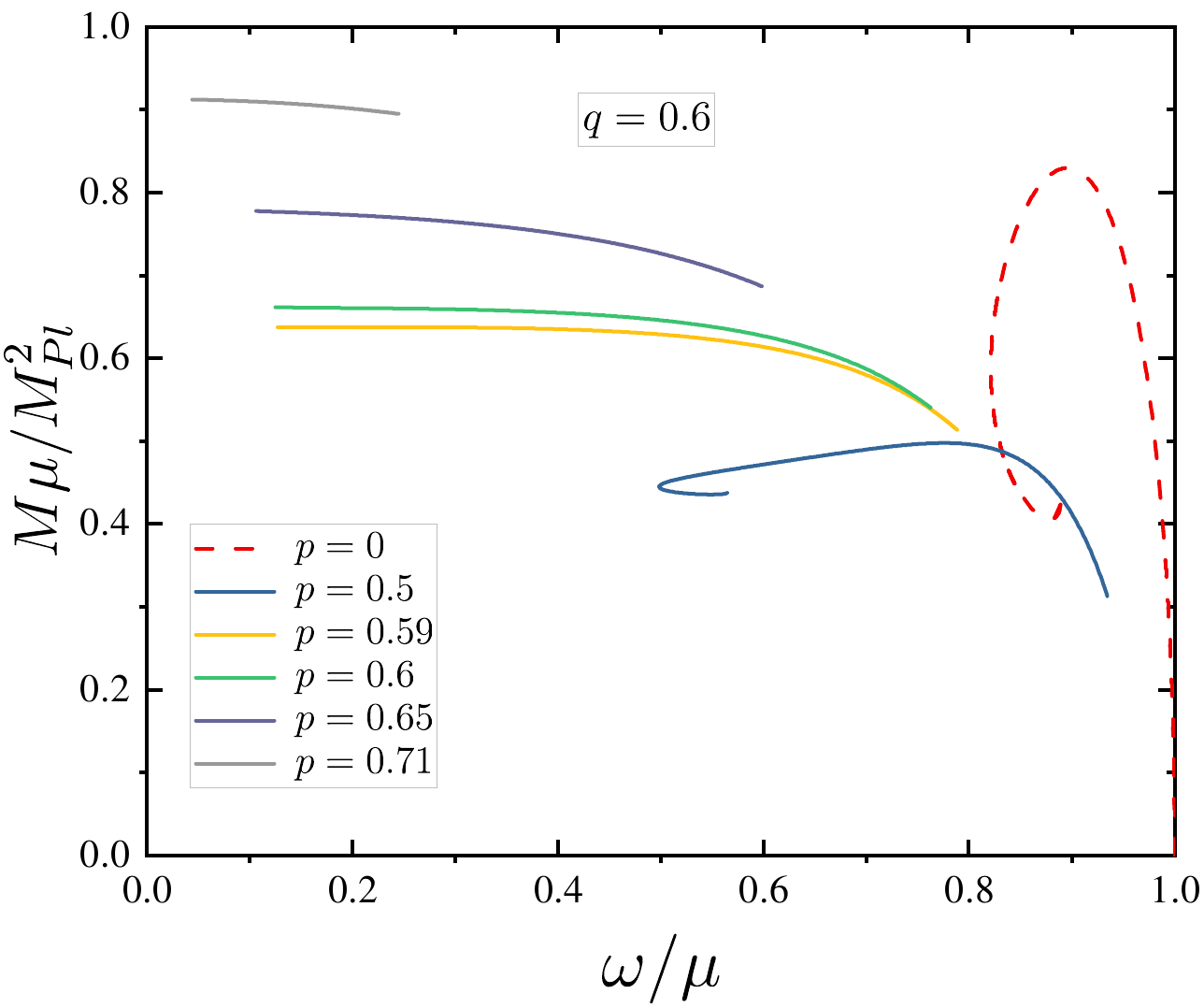}
			\label{fig:cbbsmatterfor02andall06}
		}
    		\subfigure{  
			\includegraphics[height=.28\textheight,width=.30\textheight, angle =0]{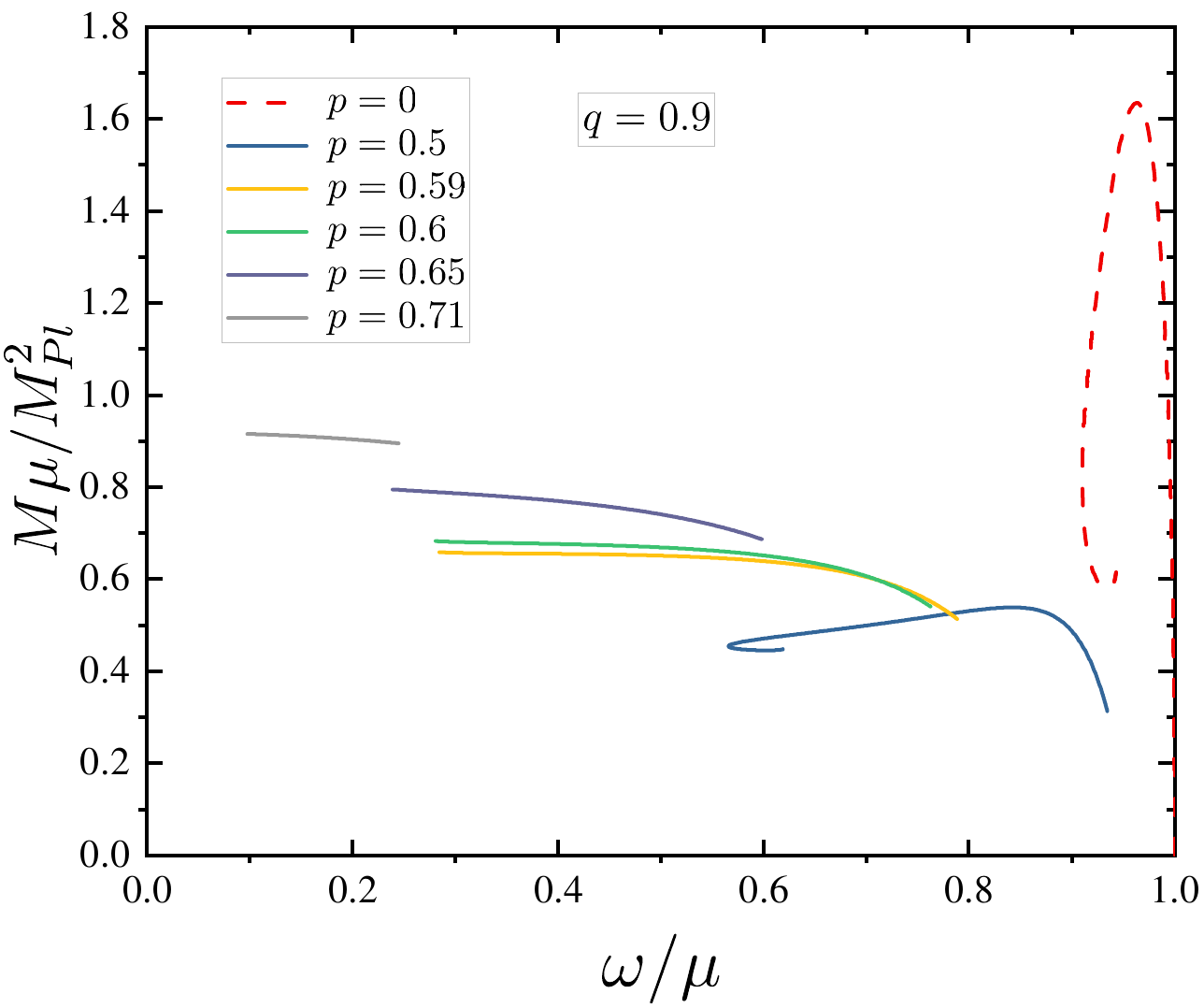}
			\label{fig:cbbsmatterfor02andall09}
		}
  		\end{center}	
		\caption{ The ADM mass $M$ of the small $q$ solution as a function of the frequency $\omega$ for different magnetic charges $p$ with $q=0,0.3,0.6,0.9$.}
		\label{fig:smallfixqmatter}		
		\end{figure}

In addition to the properties of the field functions, some physical quantities related to the solution are also crucial, such as the ADM mass. We give the curve of ADM mass $M$ versus frequency $\omega$ of small $q$ solution with $q=0,0.3,0.6,0.9$ as an example (For simplicity, this curve is called the $M$ curve) in Fig.~\ref{fig:smallfixqmatter}. From this figure, we can see that these curves have different behaviors for different magnetic charges. As shown by the red dashed line, the blue solid line of this figure, and the yellow solid line of the top panel, these curves form a series of spiral structures when the magnetic charge is small. Starting from the right end of these curves, the mass $M$ increases and then decreases as $\omega$ decreases for the first time, and this part of the solution is often referred to as the first branch. Subsequently, after reaching the minimum value of $\omega$, the mass continues to decrease and the curve spirals back to form the second branch. We find that if we continue the calculations, more branches will appear. This situation is very similar to pure boson stars, Proca stars, and Dirac stars~\cite{Herdeiro:2017fhv}. However, when the magnetic charge $p$ exceeds a certain value, the second branch will not appear. Then there will be no spiral structure. In addition, during the process of the magnetic charge going from small to large, the left endpoint of the $M$ curve is moving towards the left end of Fig.~\ref{fig:smallfixqmatter} and the minimum frequency of the solution is slowly getting smaller. In other words, the increase in magnetic charge makes the minimum frequency of the solution become smaller. It is worth noting that in addition to the magnetic charge $p$ affecting the second branch, an increase in electric charge will also cause the second branch to disappear, as shown by the solid yellow line of the bottom panel. But in contrast to the magnetic charge $p$, an increase in electric charge causes the minimum frequency to be larger while shortening the $M$ curve.

%
%
 
\begin{figure}[!htbp]
		\begin{center}
		\subfigure{ 
			\includegraphics[height=.28\textheight,width=.30\textheight, angle =0]{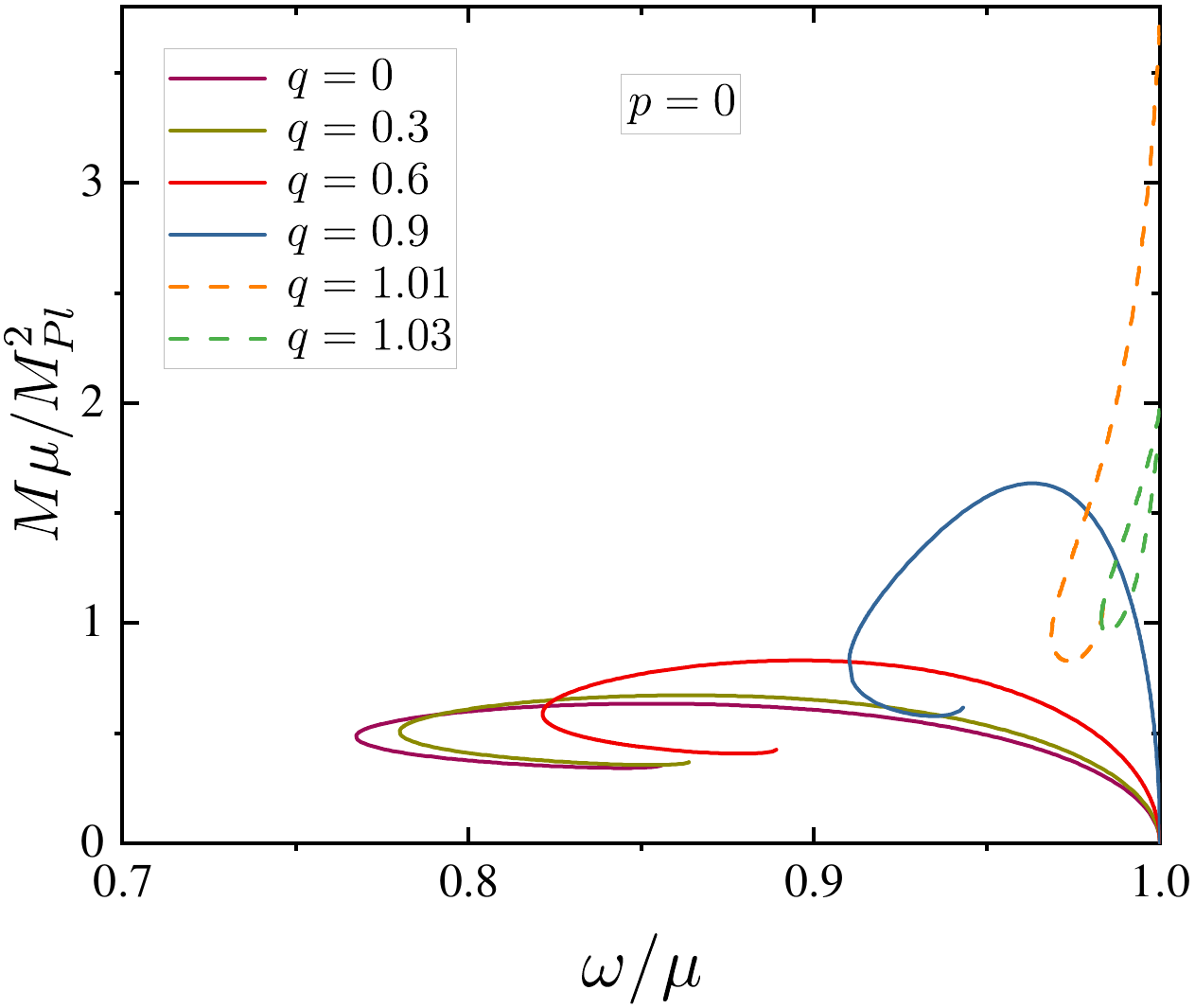}
			\label{fig:cbbsmatterfor02and00all}
		}	 
  		\subfigure{  
			\includegraphics[height=.28\textheight,width=.30\textheight, angle =0]{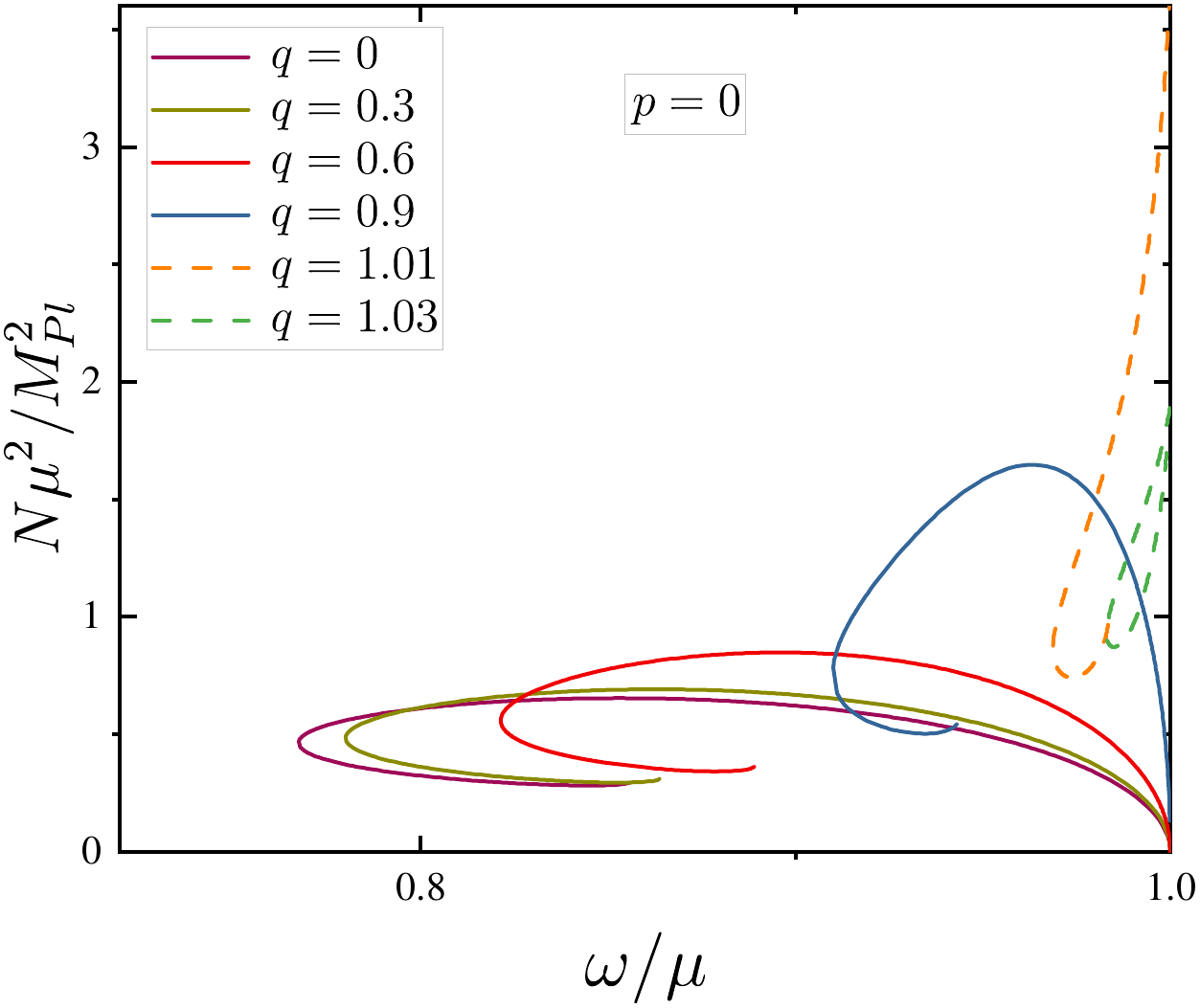}
			\label{fig:cbbsnumberfor02and00all}
		}	
  \quad
		\subfigure{ 
			\includegraphics[height=.28\textheight,width=.30\textheight, angle =0]{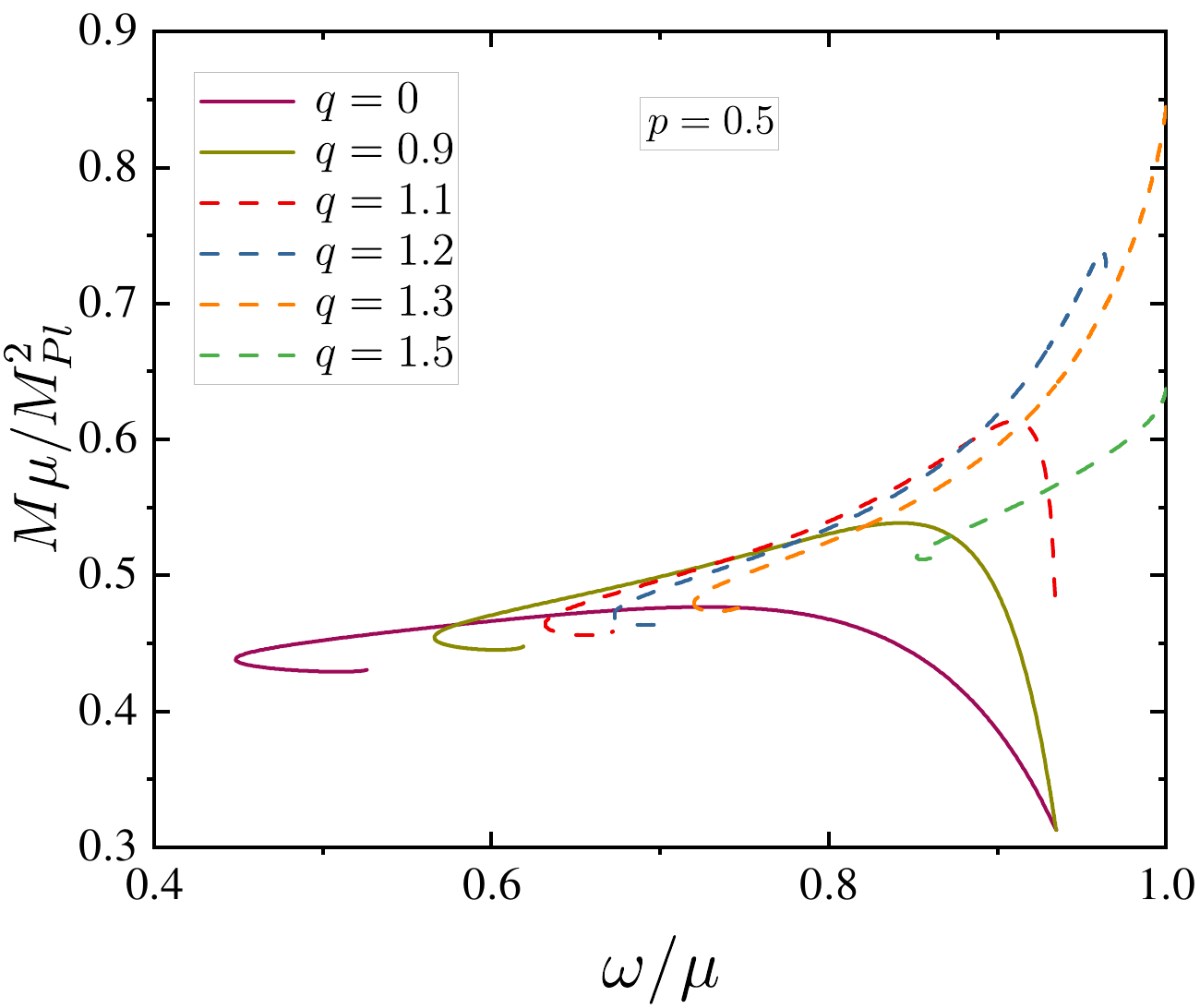}
			\label{fig:cbbsmatterfor02and05all}
		}	 
  		\subfigure{  
			\includegraphics[height=.28\textheight,width=.30\textheight, angle =0]{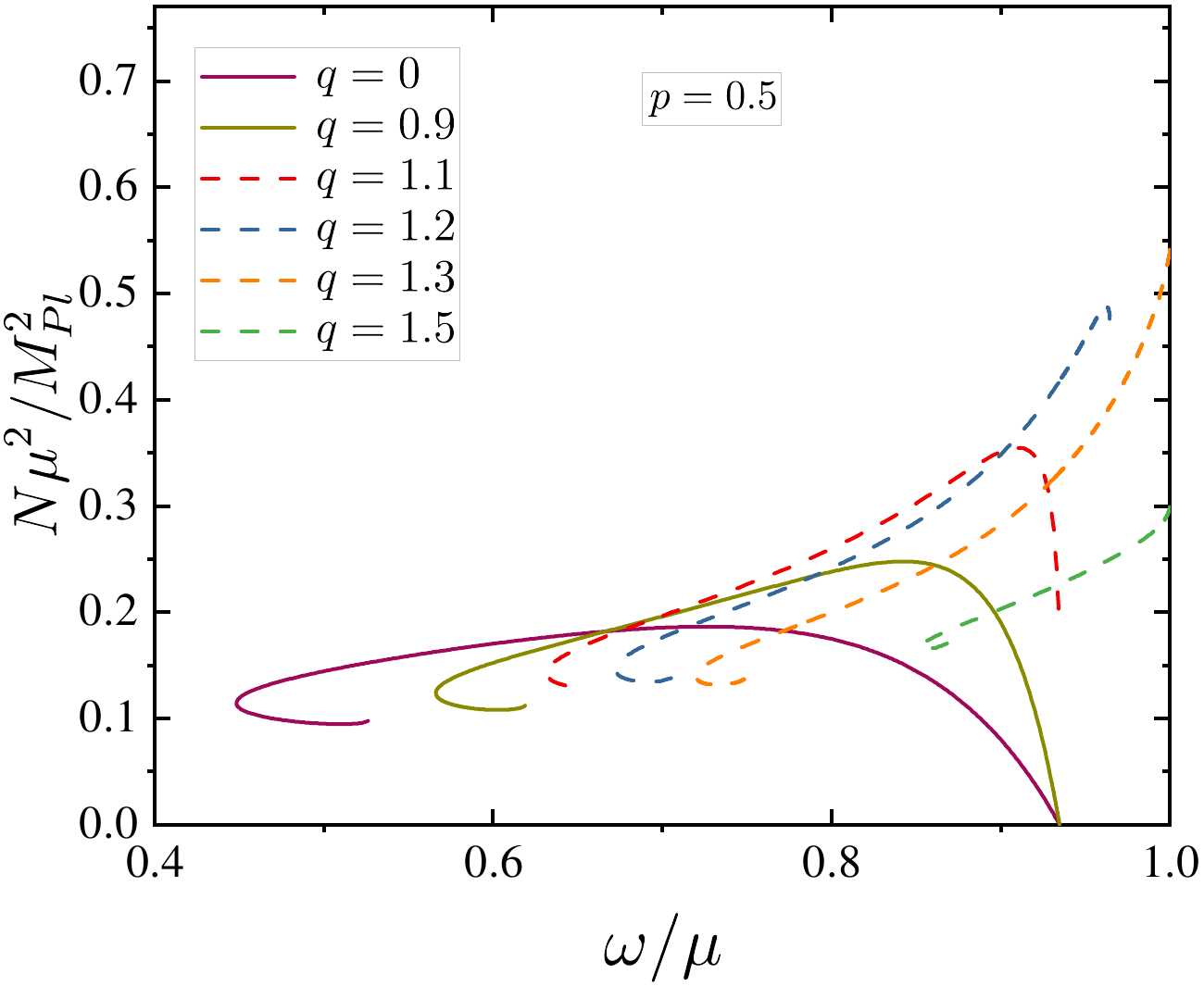}
			\label{fig:cbbsnumberfor02and05all}
		}
  \quad
  		\subfigure{ 
			\includegraphics[height=.28\textheight,width=.30\textheight, angle =0]{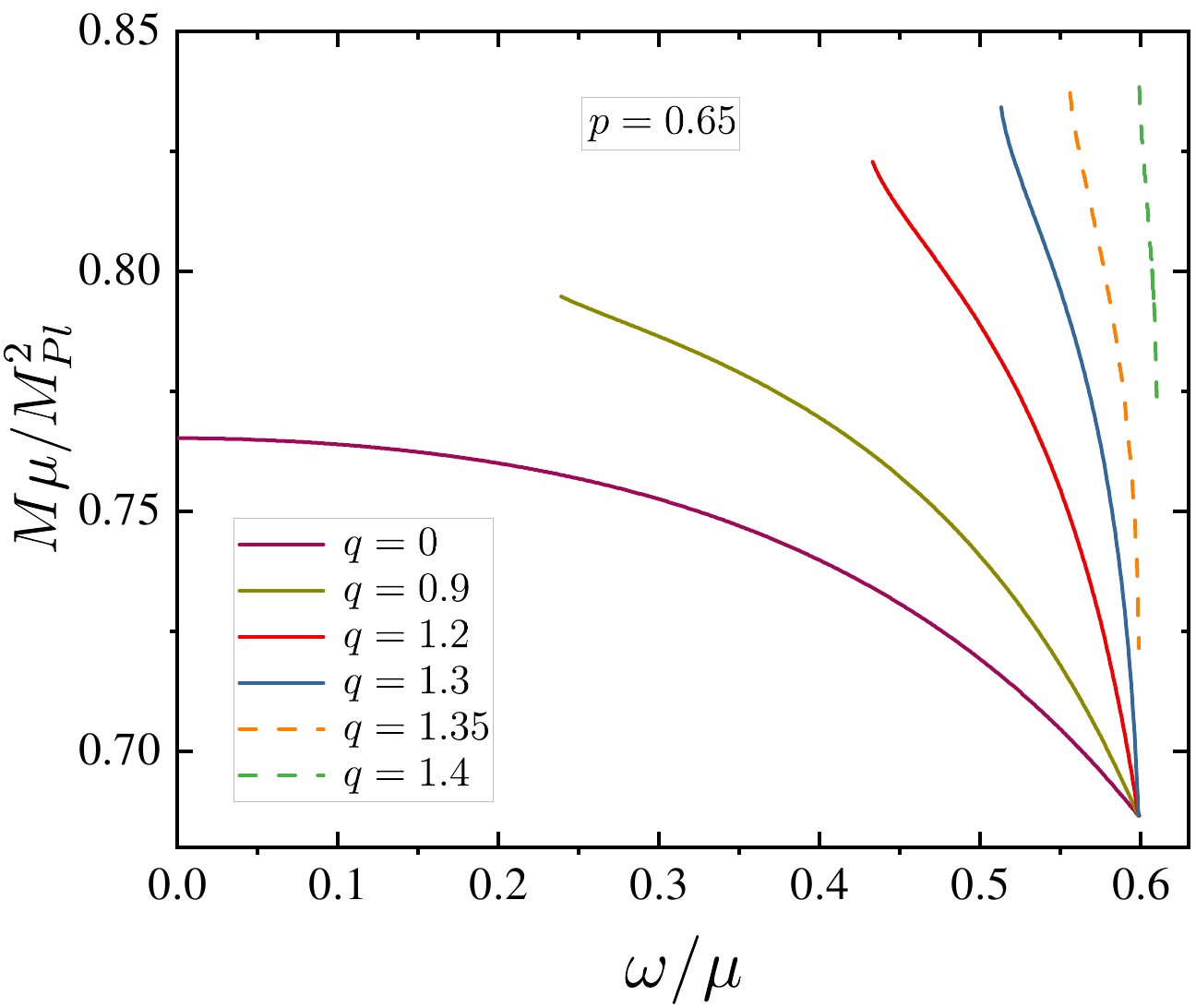}
			\label{fig:cbbsmatterfor02and065all}
		}	 
  		\subfigure{  
			\includegraphics[height=.28\textheight,width=.30\textheight, angle =0]{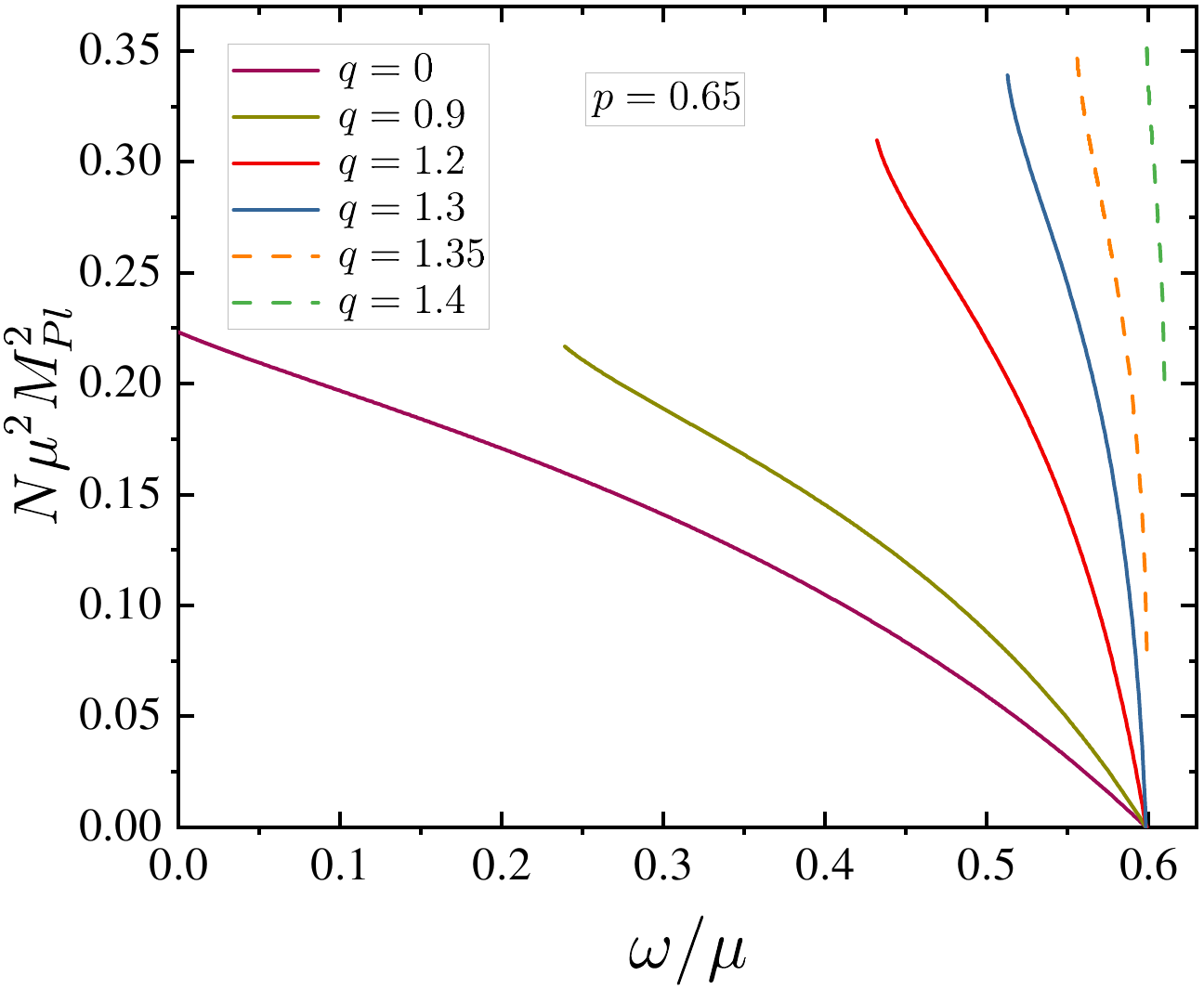}
			\label{fig:cbbsnumberfor02and065all}
		}
  		\end{center}	
		\caption{ The ADM mass $M$ (left) and the particle number $N$ (right) $vs.$ the frequency $\omega$ for the small $q$ solution.}
		\label{fig:smallfixpmatter}		
		\end{figure}

Moreover, we also found that when the electric charge is below a certain value $q_c$, under the same magnetic charge, as the charge increases, the $M$ corresponding to the maximum frequency of the small $q$ solution remains the same. Nevertheless, when the charge exceeds $q_c$, the $M$ corresponding to the maximum frequency will change as the electric charge changes, and the $M$ curve of the small $q$ solution undergoes significant changes. For the case of $p = 0$, $q_c$ refers as the ``critical charge"~\cite{Pugliese:2013gsa,Lopez:2023phk,Brihaye:2023hwg}. In this paper, we also use this term for other magnetic charges. We present the values of critical charges with several magnetic charges in Tab.~\ref{tab:rangeofq}. To better study this property, unlike Fig.~\ref{fig:smallfixqmatter}, we fixed a magnetic charge $p$ and showed the changes of the curves of the ADM mass $M$ and particle number $N$ versus frequency $\omega$ of the small $q$ solution for different magnetic charges in Fig.~\ref{fig:smallfixpmatter} (For simplicity, similar to the $M$ curve, the curve about the number of particles is called ``the $N$ curve"). 
  
First, it can be seen that from Fig.~\ref{fig:smallfixpmatter} for the same magnetic charge, as the electric charge increases, both the first branch and the second branch become shorter. This means that the existence domain of the solution with respect to frequency decreases as the electric charge increases. Second, we find that the right endpoints of the $M$/$N$ curve of the small $q$ solutions with different electric charges are at the same point if the charge is below some value $q_c$. And when the electric charge surpasses the critical charge, neither the $M$ curve nor the $N$ curve ends at the same point on the right, and the form of the $M$/$N$ curve curves also changes significantly - see the dashed line of Fig.~\ref{fig:smallfixpmatter}. Particularly, for the case of $p=0$, as shown by the dashed and solid lines respectively, the frequency corresponding to the right endpoints of the curves for the small $q$ solutions below and above the critical charge is the same, differing only in mass or particle number. As the magnetic charge increases slightly, i.e., $p\neq0$ (the middle panel and bottom panel), both the frequency, mass, and particle number corresponding to the right endpoints of the curves of the small $q$ solution are different.

 As shown in Fig.~\ref{fig:funmatter6565} and Fig.~\ref{fig:funcmetirc6565}, since the minimum of the metric function decreases with decreasing frequency, the left endpoint of the $M$/$N$ curve corresponds to the frequency where the frozen star solution is most likely to occur. Therefore, in order to better analyze the nature of the frozen star, we plot the metric function $-g_{tt}$ corresponding to the minimum frequency of the small $q$ solution (i.e., the left endpoint of the $M$/$N$ curve) for several magnetic charges in Fig.~\ref{fig:funcnochange}. From the top left panel, we can see that the minimum value $-g_{tt(min)}$ of the metric function $-g_{tt}$ at the left endpoint decreases as the electric charge increases for smaller magnetic charges. However, the top right panel shows that when the magnetic charge is a little larger, $-g_{tt(min)}$ becomes smaller and then larger as the charge increases. In particular, without $\omega\rightarrow 0$, the small $q$ solution corresponding to the solid yellow line can be regarded as a frozen star ($-g_{tt(min)}<10^{-5}$), which is impossible when the electric charge does not exist. Finally, as seen in the bottom panel, when the magnetic charge is larger, $-g_{tt(min)}$ only increases as the electric charge increases, and when the electric charge increases to a certain value, the solution can no longer be considered a frozen star (see $q=0.9$, $1.2$, $1.3$ for illustrative cases). For these solutions that cannot be considered as frozen stars, since $-g_{tt(min)}$ becomes larger as their frequency increases, these solutions cannot form frozen star solutions regardless of how their frequency changes. In other words, when the magnetic charge is large, the introduction of the electromagnetic field prevents the formation of frozen star solutions.

\subsection{Large $q$ solution (probe limit solution)}
When the electric charge $q$ is relatively large, we discover another type of solution, which we refer to as the ``large $q$ solution"\footnote{In our results, if $p=0$, the large $q$ solution does not exist.}. Unlike the small $q$ solution, this type of solution allows $q$ to tend towards infinity, and as $q\rightarrow \infty$, this solution becomes the pure Bardeen solution. Therefore, we call this solution also the probe limit solution. We can see this through the following scaling transformations 
\begin{equation}
    A \rightarrow A/q,  \quad \phi \rightarrow \phi/q.
\end{equation}
Thus the action (\ref{eq:action}) changes into
\begin{equation}
     S=\int d^4 x \sqrt{-g}\left [\frac{R}{16\pi G}+\mathcal{L}^{(1)}\right]+\frac{1}{q}\left( \mathcal{L}^{(2)}+\mathcal{L}^{(3)}\right ),
         \label{eq:action2}
\end{equation}
From the above, when $q \rightarrow \infty$ we can see that the action (\ref{eq:action}) changes to Bardeen action, and charged scalar fields cannot affect Bardeen spacetime.

	\begin{figure}[!htbp]
		\begin{center}
		\subfigure{ 
			\includegraphics[height=.28\textheight,width=.30\textheight, angle =0]{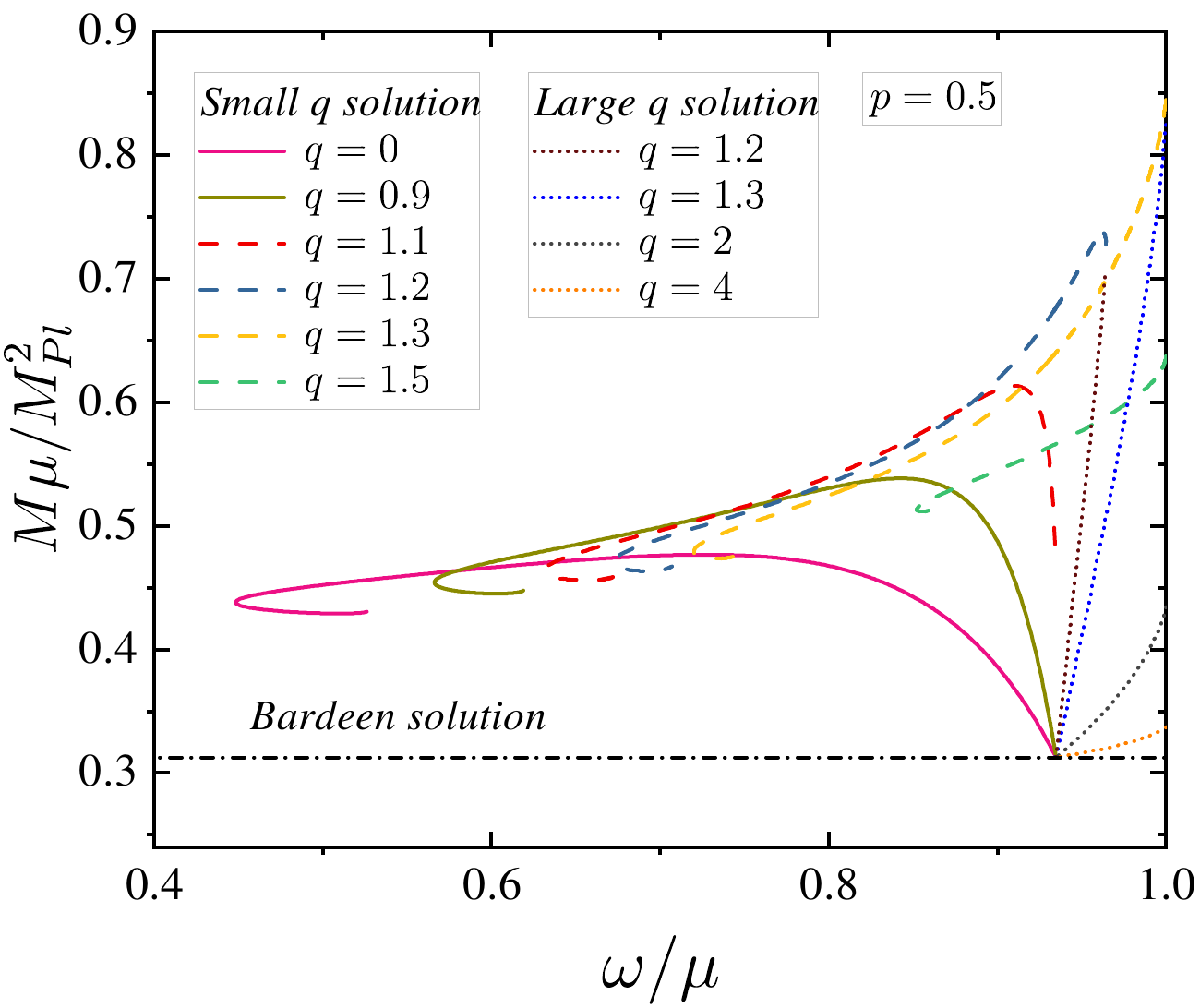}
			\label{fig:cbbsmatterfor02and05allsolution}
		}	 
  		\subfigure{  
			\includegraphics[height=.28\textheight,width=.30\textheight, angle =0]{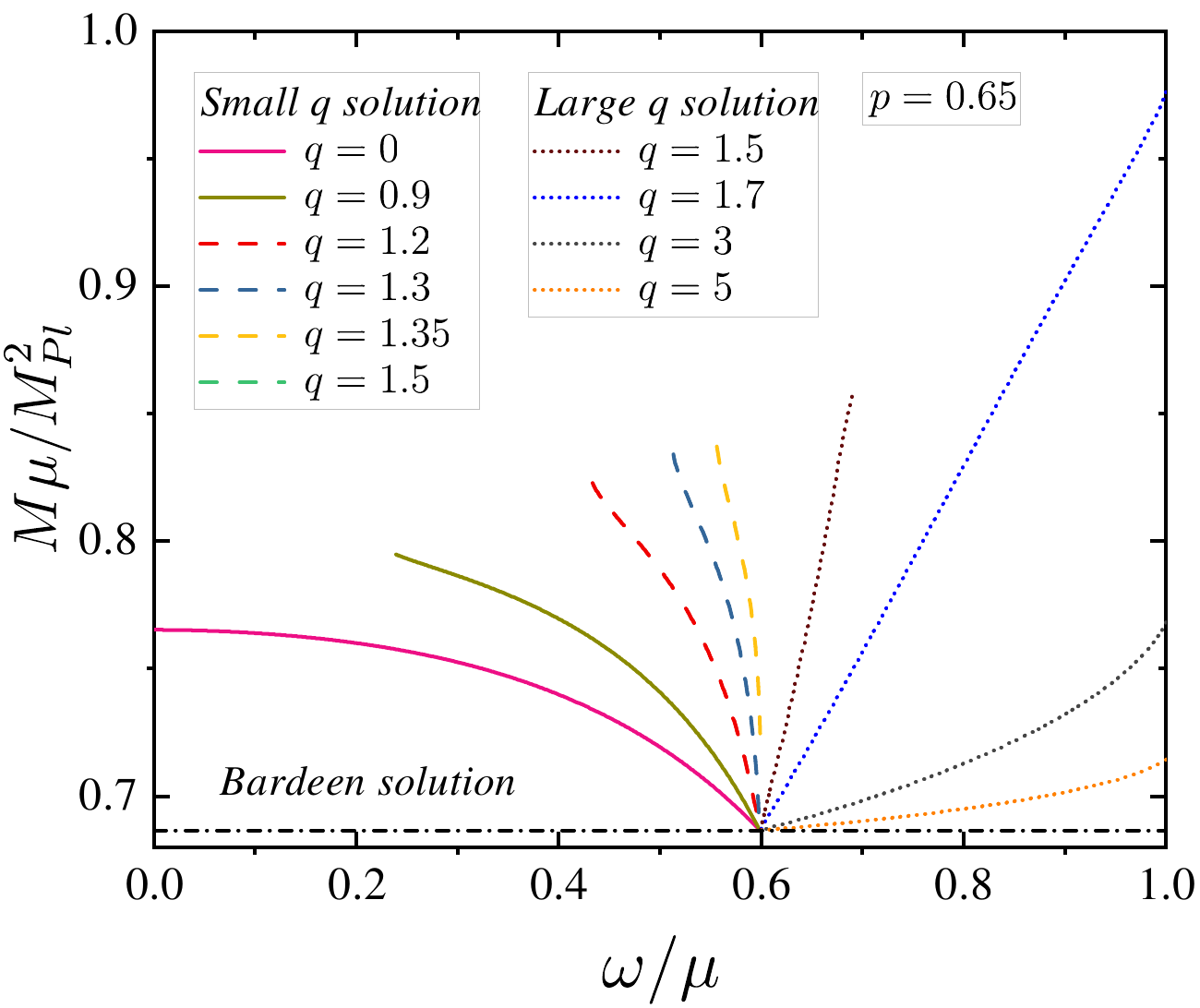}
			\label{fig:cbbsmatterfor02and065allsolution2}
		}	
  		\end{center}	
		\caption{The ADM mass $M$ as a function of the frequency $\omega$ with $p=0.5$ (left panel) and $p=0.65$ (right panel).}
		\label{fig:cbbsmatterallsolution}		
		\end{figure}

Fig. \ref{fig:cbbsmatterallsolution} shows the large $q$ solution with two magnetic charges as the illustrative case. The dotted line in the figure represents the large $q$ solution, and for illustrative purposes, we used the solid and dashed lines to represent the small $q$ solution. As shown in the figure, on the one hand, it can be observed that the left endpoint of the $M$ curve for the large $q$ solution (the dotted line) coincides with the right endpoint of the $M$ curve for the small $q$ solution below the critical charge (the solid line). This coinciding right endpoint corresponds to the pure Bardeen solution without the charged scalar field and also represents the minimum frequency for the large $q$ solution. This implies that the frequency of the large $q$ solution is not less than that of the small $q$ solution that is below $q_c$. However, on the other hand, as seen in the dashed line in this figure, the frequencies of small $q$ solutions above the critical charge do not exhibit this relationship with the frequencies of large $q$ solutions. Their frequencies usually partially overlap with those of the large $q$ solutions, and some of the small $q$ solutions that exceed the critical charge will have higher frequencies than the large $q$ solutions.

In addition, for the large $q$ solution (the dotted line), it also can be seen from Fig. \ref{fig:cbbsmatterallsolution} that when $q$ is small, the frequency cannot reach one and the ADM curve of the large $q$ solution is steep. However, as $q$ gradually increases, the domain of existence of the solution with respect to the frequency becomes larger and larger until the frequency maximum is one; at the same time, the $M$ curve becomes flatter and flatter, and finally when $q\rightarrow \infty$, the $M$ curve of the large $q$ solution coincides with the Bardeen solution (black dotted line). That is, when the charge is infinite, the solution becomes a pure Bardeen solution.


\section{CONCLUSION}\label{sec: conclusion}
In this paper, we investigated the model of Einstein-Bardeen theory coupled to a charged scalar field and obtained two types of solutions which we call the ``small $q$ solution” and the ``large $q$ solution”. There exists a limit to the magnetic charge of both solutions and within the limit, we did not find black hole solutions. 

For the small $q$ solution, there exists an upper limit to its electric charge. Within this upper limit, there exists a critical charge, and for the same magnetic charge, the curves of the ADM mass $M$ or the number of particles $N$ versus the frequency $\omega$ of the small $q$ solution above the critical charge will be significantly different from those of the small $q$ solution below the critical charge. In addition, we find that after introducing the electromagnetic field, It is possible to obtain the solution for the frozen star without needing $\omega\rightarrow 0$. However, it should be noted that this is only the case when the magnetic charge is relatively small. When the magnetic charge exceeds a certain range, the introduction of the electromagnetic field may result in solutions that fail to form frozen stars.

Instead, for the large $q$ solution, we find that there is no upper limit to its electric charge. In other words, the electric charge of the large $q$ solution can tend towards infinity. And when the electric charge tend to infinity ($q\rightarrow\infty$), the large $q$ solution becomes the pure Bardeen solution. In other words, this solution effectively serves as the probe limit solution. From the curve of the ADM mass $M$ versus the frequency $\omega$, we observe that the minimum frequency of this solution corresponds precisely to the maximum frequency of the small $q$ solution below the critical charge.

There are many interesting extensions of our work. Firstly, this work considers only the ground state solution where the charge scalar field function has no node. In our future work, we plan to consider the excited states solution with more nodes. Secondly, it is also interesting to extend our study to other matter fields including free Dirac fields and Proca fields, etc. Finally, the stability of the small $q$ solution or the large $q$ solution remains unexplored in this paper, one would wonder whether our model can be stable against small perturbations. Therefore, the dynamic stability of the solution is also a crucial subject for further investigation.
\section*{ACKNOWLEDGEMENTS}
	This work is supported by National Key Research and Development Program of China (Grant No. 2020YFC2201503) and the National Natural Science Foundation of China (Grants No.~12275110 and No.~12047501).

\end{document}